\documentclass[showpacs, prb,twocolumn,preprintnumbers , superscriptaddress, aps]{revtex4-2}
 
\usepackage{braket}
\usepackage{color}
\usepackage{amsmath,amssymb}
\usepackage{pifont}
\usepackage{amssymb}  % More symbols
\usepackage{bbold}
\usepackage{float}
\usepackage{stfloats}
\usepackage{amsmath,amssymb}
\usepackage{mathtools}
\usepackage{physics}
\usepackage{breqn} 

\usepackage[caption=false]{subfig}
\usepackage{tikz}
\usepackage{makecell}
\usepackage{pifont}   % Ding symbols
\usepackage{graphicx} % Include figure files
\usepackage{dcolumn}  % Align table columns on decimal point
\usepackage{bm}       % bold math
\usepackage{multirow} % Table functions
\usepackage{placeins}% For making floats not move around everywhere
\usepackage[colorlinks]{hyperref}
\usepackage{mathtools}

\renewcommand{\vec}[1]{\mathbf{#1}}

\usepackage{breqn}
\usepackage{mathrsfs}

\usepackage{multirow}

\begin{document}

\title{Magnon bandstructure and topology in a periodically deformed Kagome lattice with DM interaction}

    \author{Mohammed Abdullah Hammadi}
  \affiliation{Physics Department$,$
  King Fahd University
  of Petroleum $\&$ Minerals$,$
  Dhahran 31261$,$ Saudi Arabia}
    \author{Mohammed Salman Alsadah}
  \affiliation{School of Physics and Astronomy$,$ The University of Edinburgh}
 \author{Husam Abdulmajeed Noorwli}
  \affiliation{Physics Department$,$
  King Fahd University
  of Petroleum $\&$ Minerals$,$
  Dhahran 31261$,$ Saudi Arabia}
    \author{Hocine Bahlouli}
  \affiliation{Physics Department$,$
  King Fahd University
  of Petroleum $\&$ Minerals$,$
  Dhahran 31261$,$ Saudi Arabia}
    \author{Michael Vogl}
  \affiliation{Physics Department$,$
  King Fahd University
  of Petroleum $\&$ Minerals$,$
  Dhahran 31261$,$ Saudi Arabia}
      \affiliation{Interdisciplinary Research Center (IRC) for Advanced Quantum Computing (AQC) $,$ KFUPM$,$ Dhahran$,$ Saudi Arabia}

\begin{abstract}
We study the band structure and topology of magnons for a Heisenberg model with Dzyaloshinskii–Moriya (DM) interaction on a deformed Kagome lattice. For simplicity, we focus on a periodically deformed lattice with hexagonal symmetry and an enlarged unit cell. This enlarged unit cell gives rise to a more complex band structure than in the undeformed case. Analyzing band topology, there is a distinction between the topologically trivial case with anti-ferromagnetic coupling and the topologically non-trivial case with ferromagnetic coupling. In the anti-ferromagnetic case, a spin-space symmetry, also present in the classical ground state, enforces these topologically trivial states. In the ferromagnetic case, this symmetry is spontaneously broken by the classical ground state. Consequently, the band structure also hosts numerous topological features. Specifically, we observe many topological transitions and bands with Chern numbers ranging from $+2$ to $-2$, which makes the system richer than in the undeformed case. This emergent topological structure demonstrates that deformed magnets can host new and exciting physics. 
\end{abstract}

\maketitle

\section{Introduction}
Magnetism is one of the most extensively studied areas of theoretical physics because of both its rich phenomenology and its wide range of technological applications. At its core, magnetism is an inherently quantum-mechanical phenomenon, arising from the spin degrees of freedom of constituent particles and the quantum nature of their interactions. One important feature of magnetism is the emergence of magnetic order from the collective dynamics of interacting spins. The theoretical description of magnetism in quantum systems began with the introduction of the Heisenberg model in 1928 \cite{Heisenberg1928}. Soon thereafter, Felix Bloch \cite{bloch1930theorie} introduced the concept of spin waves, which represent collective excitations of localized spins about their ordered ground state, thereby providing a microscopic understanding of ferromagnetism.

In addition to their diverse theoretical properties, magnetic materials play a crucial role in modern technology. A prominent example is the read-head sensor in hard disk drives, which consists of alternating ferromagnetic and nonmagnetic conducting layers and operates via the so-called giant magnetoresistance (GMR) effect. In these systems, spin-dependent scattering gives rise to large changes in electrical resistance \cite{baibich1988giant,binasch1989enhanced}. More advanced implementations exploit tunnel magnetoresistance (TMR) \cite{miyazaki1995giant,yuasa2004giant,moodera1995large,yuasa2007giant, parkin2004giant}, which is particularly suitable for high-density data storage and magnetic random-access memory (MRAM)\cite{bhatti2017spintronics, nakatani2018read, dieny1994giant}. These examples demonstrate the versatility of magnetic systems and their capacity to exhibit technologically useful phenomena.

This success of spintronic technologies motivates the search for exotic magnetic states with novel functionalities. In this context, frustrated lattices, such as the Kagome lattice, have attracted considerable attention as promising platforms for realizing quantum spin liquid phases, in which spin disorder persists even at absolute zero temperature \cite{savary2017quantum}. Remarkably, certain spin liquid states are predicted to support non-Abelian anyons - emergent quasiparticles whose quantum states are topologically protected against local perturbations and decoherence, thereby offering promising prospects for robust quantum information processing.\cite{kitaev2006anyons,nayak2008non}. Topological phenomena in magnetic systems, however, are not restricted to quantum spin liquids and can also arise in magnetically ordered phases through the topology of spin textures and collective excitations.

Such topological phases of matter lie beyond the scope of Landau’s theory of phase transitions \cite{landau1937theory}, which characterizes phases in terms of symmetry breaking. Specifically, certain quantum phases, including chiral spin states and quantum spin liquids, cannot be characterized within Landau’s symmetry-breaking framework. Their description instead requires the concept of \textit{topological order} \cite{wen1990topological}, which is often associated with long-range quantum entanglement. Such phases can instead be characterized by topological invariants, such as the Chern number \cite{thouless1982quantized}. A prominent example is the quantum Hall effect \cite{klitzing1980new}, where the Hall conductance is quantized and directly determined by the Chern number.

In magnetic systems, topological properties can emerge in both real-space spin textures and the band structure of collective excitations. One prominent example is a magnetic skyrmion, a topologically nontrivial spin texture characterized by a winding number, which, owing to its topological protection, is robust against thermal fluctuations and thus attractive for information storage and processing \cite{fert2017magnetic}.  A second example arises in the context of topological quantum computing, where non-Abelian anyons, emerging from quantum spin liquid phases, can serve as intrinsically robust qubits \cite{nayak2008non}. These and related phenomena provide strong motivation for the continued exploration of topological effects in magnetic systems.

An important question is how to control and engineer these topological and magnetic properties through external perturbations. One particularly versatile mechanism is lattice deformation. Well-known examples of this concept appear outside the context of magnetism. For instance, in twisted bilayer graphene, a moir\'e lattice forms due to the misalignment of two identical layers, leading to flat electronic bands and the appearance of correlated phases such as unconventional superconductivity and Mott insulating states \cite{cao2018unconventional}. Similarly, lattice deformations can also give rise to exotic phases of matter. In the case of graphene, for instance, deformation generates a pseudomagnetic field \cite{guinea2010energy}, which can give rise to exotic phases of matter, such as the fractional quantum Hall effect. 

These examples highlight lattice deformations as a powerful mechanism for engineering nontrivial topology, tailored band structures, and exotic phases of matter. This naturally raises the question of whether similar phenomena can be realized in magnonic systems. Indeed, previous studies have shown that uniaxial strain applied to a ferromagnetic Kagome lattice with Dzyaloshinskii--Moriya interaction can drive topological phase transitions \cite{owerre2018strain}. More recently, higher-Chern-number magnon phases have also been obtained in Kagome ferromagnets through the interplay of Dzyaloshinskii--Moriya and pseudodipolar interactions \cite{Ni2025}. In contrast to these mechanisms, we consider a two-dimensional periodic deformation that preserves sixfold rotational symmetry while enlarging the real-space unit cell. The resulting reduced Brillouin zone and deformation-induced hybridization of folded bands provide a qualitatively different route to engineering magnon topology. The Kagome lattice, already known for its nontrivial topological properties, provides an ideal platform for this investigation. Our choice of deformation preserves the fundamental lattice structure while enabling access to previously unexplored regimes of magnon band topology. The primary objective of this work is to elucidate the interplay between magnons and periodic lattice deformations within a minimal theoretical framework.

The remainder of the paper is organized as follows. In Sec. \ref{sec:IIrev}, we begin by studying the undeformed Kagome lattice, which serves as a baseline for the deformed case. Then, in Sec. \ref{sec:IIIdeflatt}, we introduce the mathematical form of the deformation via a vector field and examine its effect on the classical ground-state configuration and the resulting band structure. Subsequently, in Sec. \ref{sec:IVbandtopo}, we study the topological properties of the bands. Finally, in Sec. \ref{sec:conclusion} we present our conclusions.

\section{Review of undeformed Model and simplified solution}
\label{sec:IIrev}

Our goal in this paper is to study the properties of a Kagome lattice with DM interaction after it is subjected to a hexagonal deformation field. Before discussing the deformed model, we review the undeformed case to keep our work self-contained and provide a reference point for comparison.
    \begin{figure}[H]
    \centering
\includegraphics[width=\linewidth]{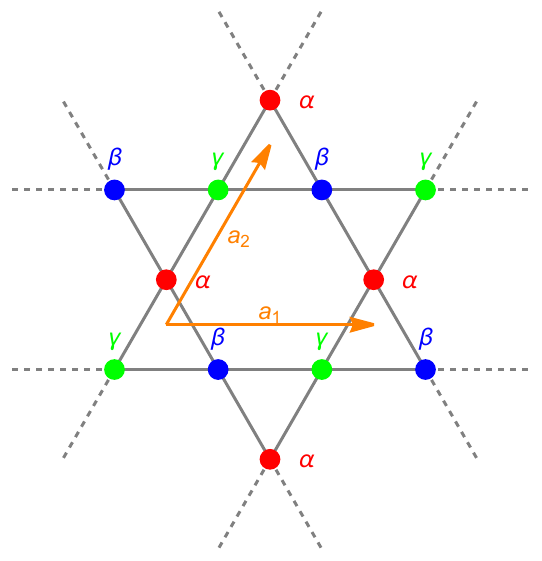}
    \caption{Kagome lattice with its three inequivalent sites $\alpha$,$\beta$,$\gamma$ labeled.}
    \label{Kagome}
    \end{figure}
A typical Kagome lattice is shown in Fig. \ref{Kagome}.
As shown, the Kagome lattice has two primitive unit cell vectors, $\vec a_1$ and $\vec a_2$, forming a parallelogram-like unit cell. At each lattice point, the basis consists of three inequivalent sites that we label $\alpha$, $\beta$, and $\gamma$. Therefore, the Kagome lattice is not a Bravais lattice by itself. Rather, its unit cells form a triangular lattice with three sublattices. Positions of unit-cell centers are then given by
    \begin{equation}
   \mathbf{R} = n_1 \mathbf{a}_1+n_2\mathbf{a}_2
\end{equation}
with
\begin{equation}
    \mathbf{a}_1=(2,0), \hspace{1cm} \mathbf{a}_2=(1,\sqrt{3})
\end{equation}
as basis vectors. Different sublattices can be accessed by shifts
\begin{equation}
 \boldsymbol\delta_\alpha=\left(0, \frac{1}{\sqrt{3}}\right), \quad \boldsymbol\delta_\beta=\left(\frac{1}{2},-\frac{\sqrt{3}}{6}\right), \quad \boldsymbol\delta_\gamma=-\left(\frac{1}{2},\frac{\sqrt{3}}{6}\right).
\end{equation}

The Hamiltonian that describes interactions between the spins on the different sites is given as
    \begin{equation}
    H_{H+DM} =  \sum_{\langle i,j \rangle} \left[ J_{ij} (\mathbf{{S}}_i \cdot \mathbf{{S}}_j) + \mathbf{{D}}_{ij} \cdot (\mathbf{{S}}_i \times \mathbf{{S}}_j) \right].
\end{equation}
 Here and throughout this work, $\langle i, j\rangle$ denotes nearest-neighbor pairs. We restrict ourselves to nearestneighbor Heisenberg and DM interactions in order to isolate the effects of periodic deformation within a minimal model. The first term inside the sum is the so-called Heisenberg interaction with its corresponding exchange interaction strength $J_{ij}$. This term for the spins $\vec {S} _ {i} $ and $\vec {S} _ {j} $ favors either alignment or anti-alignment. We call the case $J_{ij}>0$ anti-ferromagnetic, where the spins favor anti-alignment. In contrast, the case $J_{ij}<0$ is called ferromagnetic and favors orderings with spins that are aligned. The second term in the Hamiltonian is called the Dzyaloshinskii–Moriya interaction (DMI)\cite{dzyaloshinsky1958thermodynamic}. This term arises due to the spin-orbit coupling and is allowed if inversion symmetry between neighboring sites is broken\cite{moriya1960anisotropic}. The DM interaction leads to spin canting. The term $\mathbf{D}_{ij}$ is called the DM vector, and determines the strength of the DM interaction and the preferred canting orientation. Its direction is fixed by the symmetry of the lattice and geometry of bonds \cite{moriya1960anisotropic}.

As we will see, it is possible to find a classical ordering for a  magnetic ground state. This observation is a significant simplification, as it allows linear spin-wave theory to serve as an approximate description of magnons \cite{holstein1940field}. Here, small fluctuations of the spins around the $z$-axis correspond to low-energy excitations that we call magnons.  We note in passing that a spin-specific related expansion that could be applied to non-ordered phases is given in \cite{PhysRevResearch.2.043243,10.21468/SciPostPhys.10.1.007,c3n8-1h7f}. To apply linear spin wave theory and obtain the energies of the magnons, we will use the Holstein-Primakoff transformation up to first order. The transformation is given by
\begin{equation}
    S^+=\hbar \sqrt{2S}\sqrt{1-\frac{a^\dagger a}{2S}}\,a \approx \sqrt{2S}\,a;\quad S_z =\hbar (S-a^\dagger a)
    \label{eq:Holstein-primakoff}
\end{equation}

We stress that this approximation is only valid if all spins in the classical ground state are oriented along the $z$- direction. So after we determine a classical ground state, a unitary transformation has to be applied in spin space that rotates all spins to align with the positive $z$-axis.

\begin{figure}[htb]
\centering
\includegraphics[width=7cm]{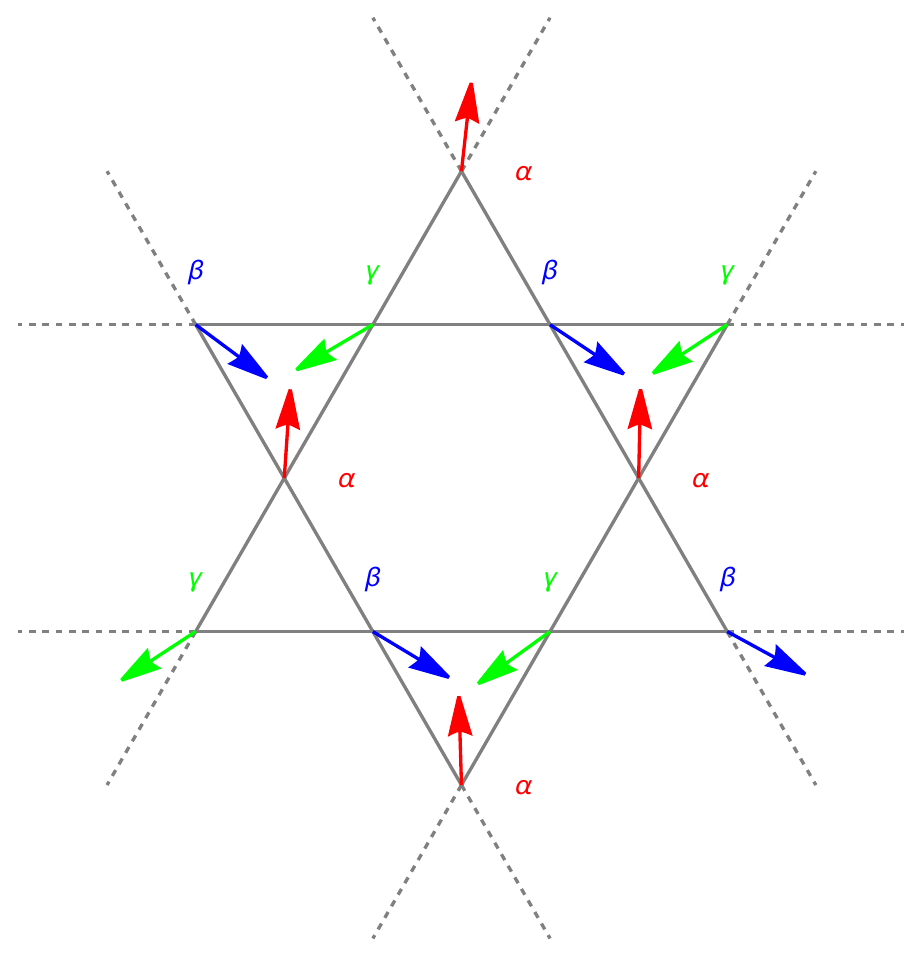}
    \caption{Classical ground state configuration of the anti-ferromagnetic case. The spin directions are $120^\circ$ separated} 
    \label{spin}
\end{figure}

We next determine the classical ground state for the antiferromagnetic case, as illustrated in Fig.\ref{spin}.

To determine this ground state, we first performed numerical calculations to identify the orientations of the classical spins. Specifically, we considered a finite lattice of size $N_x = 20$ and $N_y = 20$ unit cells in order to avoid making a priori assumptions about the magnetic unit-cell size. We then verified that spins near the center of the lattice exhibit a periodic pattern, thereby minimizing finite-size effects. An important effect to note is that symmetry under rotation around the $z$-axis implies a ground-state degeneracy under global rotations around $z$. We chose the specific orientation of spins shown in Fig. \ref{spin}. Here, spins are separated from each other by a $120^\circ$ relative angle.
The classical spin directions for the ground state were then found by working with a 3-site unit cell (sites $\alpha$, $\beta$, and $\gamma$), and their expressions are given as
 \begin{equation}
    \begin{aligned}
        &\mathbf{S}_\alpha = \left( 0, 1, 0 \right);\quad \mathbf{S}_\beta = \left( \frac{\sqrt{3}}{2}, -\frac{1}{2}, 0 \right);\quad 
\mathbf{S}_\gamma =- \left( \frac{\sqrt{3}}{2}, \frac{1}{2}, 0 \right).
    \end{aligned}
    \end{equation}
    and the corresponding ground state energy is given as ( we set spin $S=1$ and keep it for all that follows)
\begin{equation}
    E_{ground} = -3N(J+\sqrt{3}D_z).
\end{equation}

We can now write the Hamiltonian in a frame where spin is classically along the z-axis. i.e. by rotating $\tilde{\vec S} = R\vec S$
    \begin{equation}
H=\frac{1}{2}\sum_{<i, j>} \tilde {S_i^u}\Gamma_{ij}^{uv}\,  \tilde S_j^v .
\end{equation}
Here, the factor $1/2$ is to avoid double-counting. Indices $u$, and $v$ $\in \{x, y, z\}$, and matrix elements $\Gamma^{uv}_{ij}$ are defined as
\begin{equation}
\Gamma^{uv}_{ij}
= \sum_{ab}R_i^{ua}J_{ij}\,\delta^{ab}R_j^{vb}
+  \sum_{abc}R_i^{ua}\epsilon^{abc}\,D_{ij}^{c}R_j^{vb}.
\end{equation}
For our specific case, components of $\Gamma$ are explicitly given as
\begin{equation}
\begin{aligned}
    \Gamma_{ij}^{xx} &= J_{ij} \\
\Gamma_{ij}^{xy}  &= \Gamma_{ij}^{xz}  = \Gamma_{ij}^{yx}  = \Gamma_{ij}^{zx} = 0\\
\Gamma_{ij}^{yy} &= \Gamma_{ij}^{zz} = J_{ij}\cos{(\theta_i - \theta_j)} - D_{ij}^z\sin{(\theta_i - \theta_j)}  \\
\Gamma_{ij}^{zy} &= -\Gamma_{ij}^{yz} = D_{ij}^z\cos{(\theta_i - \theta_j)} + J_{ij}\sin{(\theta_i - \theta_j)} .
\end{aligned}
\end{equation}
We stress that 
$\Gamma_{ij}^{yy}$ and $\Gamma_{ij}^{zz}$ are symmetric under exchange of indices $i\leftrightarrow j$.

%We can write the Hamiltonian explicitly as a sum of unit cells as: 

%\begin{align}
%\hspace{1em}
%\space\space\space H = \sum_{\text{n1, n2}} \Big[ 
% & S^u_\alpha(n_1, n_2)\,\Gamma^{uv}_{\alpha\beta}\,S^v_\beta(n_1, n_2) \notag \\
%+& S^u_\alpha(n_1 + 1,\, n_2 - 1)\,\Gamma^{uv}_{\alpha\beta}\,S^v_\beta(n_1, n_2)  \notag \\
%+& S^u_\beta(n_1, n_2)\,\Gamma^{uv}_{\beta\gamma}\,S^v_\gamma(n_1, n_2) \notag \\
%+& S^u_\beta(n_1, n_2)\,\Gamma^{uv}_{\beta\gamma}\,S^v_\gamma(n_1 + 1, n_2) \notag \\
%+& S^u_\gamma(n_1, n_2)\,\Gamma^{uv}_{\gamma\alpha}\,S^v_\alpha(n_1, n_2)  \notag \\
%+& S^u_\gamma(n_1, n_2 + 1)\,\Gamma^{uv}_{\gamma\alpha}\,S^v_\alpha(n_1, n_2)
%\Big]. \notag
%\end{align}

After our rotation and using the Holstein-Primakoff Eq. \eqref{eq:Holstein-primakoff} transformation together with a $1/S$ expansion, we are in a position to perform a Fourier transform to block-diagonalize the Hamiltonian.
 \begin{align}
a_{i}&= \frac{1}{\sqrt{N}} \sum_{\mathbf{k}}  a_{i}(\mathbf{k}) \, e^{i \mathbf{k} \cdot \mathbf{r}_{i}}   \notag 
\end{align}

The resulting Hamiltonian is quadratic and of bosonic Bogoliubov-type, and given as
\begin{align}
H = \sum_{\mathbf{k}}
\boldsymbol{\Psi}^\dag(\vec k)h(\vec k)
\boldsymbol{\Psi}(\vec k);\quad \boldsymbol{\Psi}(\vec k)=\begin{pmatrix}
\vec a(\mathbf{k}) \\ \vec a^{\dagger}(-\mathbf{k})
\end{pmatrix}
\end{align}
where $\vec {a}(\vec k) $ is a vector of annihilation operators and different components correspond to different sublattices. We note that the expression only includes terms to order $\mathcal{O}(S^1)$. That is, while we technically set $S=1$, one can easily obtain results for other spins by renormalizing excitation energies as $E\to ES$.

The corresponding single-body Hamiltonian is given as
\begin{align}
  h(\vec k) =  \begin{pmatrix}
A(\mathbf{k}) & B(\mathbf{k}) \\
B^{*}(-\mathbf{k}) & A^{*}(-\mathbf{k}).
\end{pmatrix} 
\end{align}
Matrices $A(\vec k)$ and $B(\vec k)$ are given as
\begin{equation}
    A(\mathbf{k}) = \frac{1}{4}\sum_{\langle i , j\rangle}\Gamma_{ij}^{(+)}e^{i \mathbf{k}\cdot \mathbf{r}_{ij}}\ket{i}\otimes\bra{j}+\sum_{i}\Gamma_{ii}^{(z)}\ket{i}\otimes\bra{i}
\end{equation}
and
\begin{equation}
    B(\mathbf{k}) = \frac{1}{4}\sum_{\langle i , j\rangle}\Gamma_{ij}^{(-)}e^{i \mathbf{k}\cdot \mathbf{r}_{ij}}\ket{i}\otimes\bra{j}.
\end{equation}
We note that $i,j\in\{\alpha,\beta,\gamma\}$. For the undeformed ground-state configuration considered here, the mixed components $\Gamma_{ij}^{xy}$ and $\Gamma_{ij}^{yx}$ vanish, so the combinations entering the Bogoliubov Hamiltonian simplify to
\begin{equation}
\begin{aligned}
\Gamma_{ij}^{(\pm)}=\Gamma_{ji}^{(\pm)}= \Gamma_{ij}^{xx}\pm\Gamma_{ij}^{yy};\quad 
\Gamma_{ii}^{(z)} = \sum_{\langle i, j \rangle}-\Gamma_{ij}^{zz}.
\end{aligned}
\end{equation}

Since our Hamiltonian is a Bogoliubov-type bosonic Hamiltonian, it can be diagonalized using para-unitary transformations. For completeness and to keep our work self-contained, we give a brief description of the procedure. That is, we insert identities built from the paraunitary matrix $P$ with $P^\dag \eta P=\eta$ and $\eta=\mathrm{diag}(\mathbb{1},-\mathbb{1})$ into the Hamiltonian. Specifically we have\cite{COLPA1978327,PhysRevB.87.174427}
\begin{equation}
    \mathbb{1}=P^\dag\eta P\eta=P^{-1}P
\end{equation}
such that
\begin{equation}
    H=\sum_{\mathbf{k}}\boldsymbol{\Psi}^\dag(\vec k) P^\dag\eta P\eta h(\mathbf{k})P^{-1}P\boldsymbol{\Psi}(\vec k)
    \label{eq:para_unit_diag}
\end{equation}
Now we can choose $P$ in such a way that it diagonalizes $\eta h(\vec k)$ and we may "normalize" eigenvectors such that we can build $P^{-1}$ from eigenvectors $u_{n}(\vec k)$ with $P^\dag \eta P=\eta$ to obtain a para unitary transform. It is important to keep in mind that the procedure will only work if $H$ is positive definite\cite{COLPA1978327}, which serves as a consistency check. It is also important to note that while $\eta h(\vec k)$ has both positive and negative eigenvalues, negative eigenvalues will enter as positive terms in $H$ because of the first $\eta$ in the equation that turns them positive again. Lastly, after the transform we can immediately read off from \eqref{eq:para_unit_diag}, that we have a new Nambu spinor $\boldsymbol{\Psi}^\dag P^\dag$ and we therefore find new creation and annihilation operators are given as $b^\dag_n(\vec k)=\boldsymbol{\Psi}(\vec k)\eta \vec u_n(\vec k)$ and $b_n(\vec k)=-\boldsymbol{\Psi}(\vec k)\eta \vec u_{N+n}(\vec k)$ with $N$ the dimension of $A(\vec k)$ \footnote{It is useful to first prove that $P^{\dag}=\eta P^{-1}\eta$}.

\begin{figure}[H]
    \centering
\includegraphics[width=\linewidth]{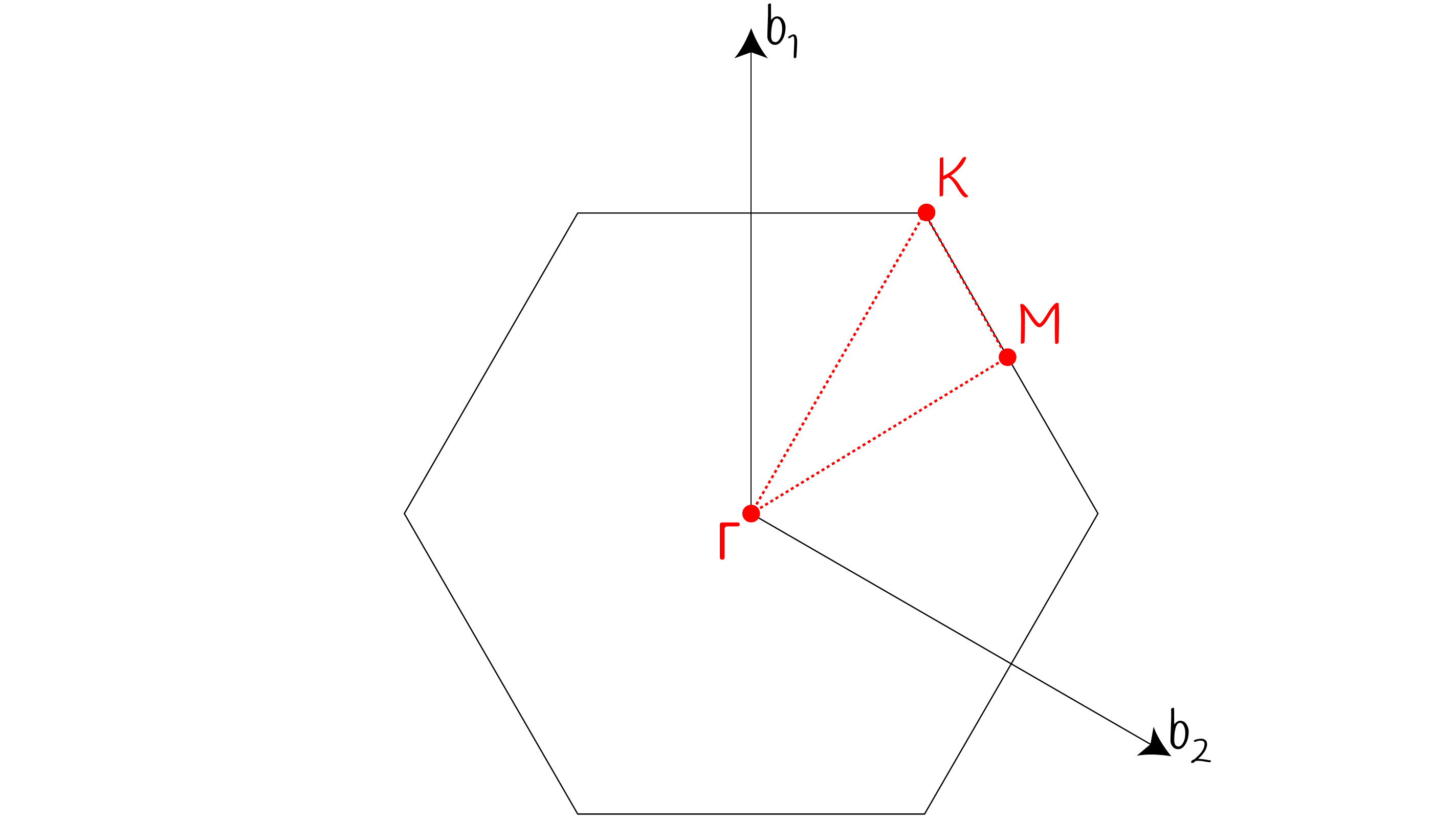}
    \caption{Brillouin zone for the undeformed Kagome lattice. $\Gamma MK\Gamma$ is a high-symmetry path used in later plots of band structures. Reciprocal lattice vectors that were used to construct high symmetry points are given as $\vec b_1 = \left( \pi, -\frac{\pi}{\sqrt{3}} \right)$ and $\vec b_2 = \left( 0, \frac{2\pi}{\sqrt{3}} \right)$}
    \label{fig:Brillouin_zone}
\end{figure}

As the last step to obtain a plot of the band structure, we also need to specify a high-symmetry path through the hexagonal Brillouin zone of the Kagome lattice, as shown in Fig. \ref{fig:Brillouin_zone}.

\begin{figure}[htb!]
    \centering
    \subfloat[]{
\includegraphics[width=\linewidth]{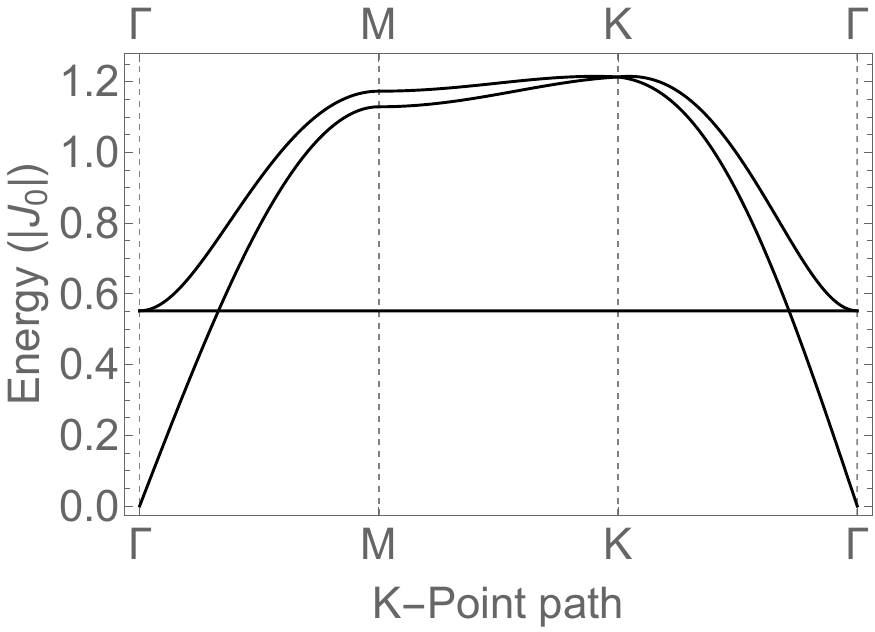}
    }
    \hfill
    \subfloat[]{
\includegraphics[width=\linewidth]{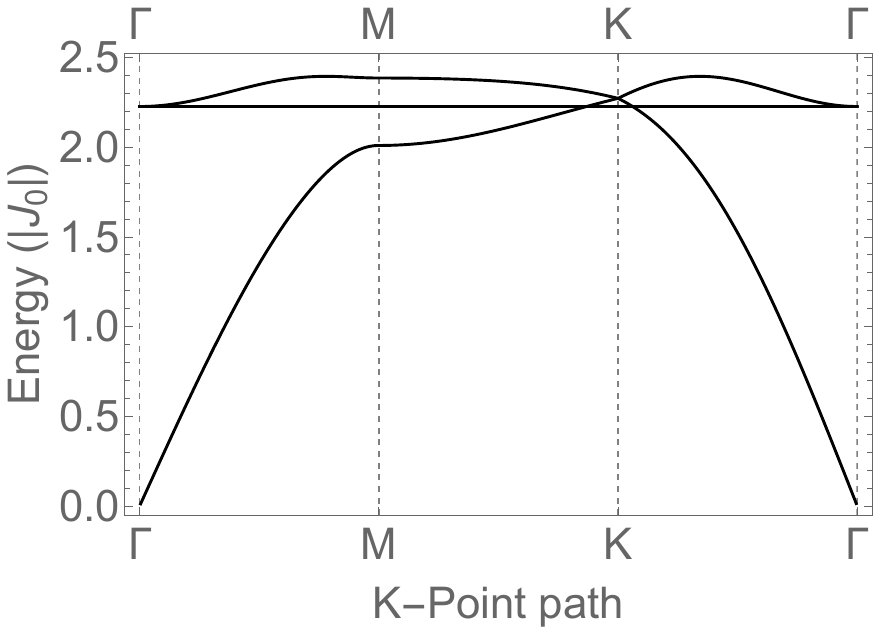}
    }
    \caption{Plot (a) shows the band structure of the undeformed Kagome lattice (antiferromagnetic case), where 
    $D_z / J$ = 0.1, while the second graph (b) shows the band structure of the same system but $D_z / J = 0.8$.}
\label{fig:band_undeformed_Kagome1111}
\end{figure}

The resulting band structure is shown in Fig \ref{fig:band_undeformed_Kagome1111}.

 Both band structures show three bands arising from the three sublattices. There are also additional copies at negative energy, which, however, don't carry meaning as additional excitation energies\cite{COLPA1978327,PhysRevB.87.174427}. Two  of the bands are dispersive, and one is globally flat. The flat band corresponds to localized (non-propagating) magnons and arises due to Geometrical frustration of the Kagome lattice, and its existence therefore does not depend on a specific ratio  $D_z/ J$. Changing the ratio $D_z/J$ also does not open any gaps, indicating that the model, as currently given in the anti-ferromagnetic regime, is ill-suited for a discussion of band topology because the Chern number is ill-defined in this case. The next step is to apply a deformation to our lattice and observe how the band structure, gaps, and possibly the topology change.

\section{Deformed lattice}
\label{sec:IIIdeflatt}
Deformations of lattices directly impact couplings between different sites. As such, it is natural to expect that they lead to new and exciting phenomena. Below, we discuss such deformations for our case.
\subsection{Choice of deformation}

When two periodic lattices with a slight mismatch are placed on top of each other, commonly moir\'e patterns appear\cite{doi:10.1073/pnas.1108174108,PhysRevB.90.155406,miftah_saud_hbngraph}. A popular example is twisted bilayer graphene, where the mismatch is in relative angle. It has been found that such seemingly small changes can lead to dramatic consequences that first manifest in a much smaller Brillouin zone, corresponding strongly altered band structures and novel phases of matter\cite{doi:10.1073/pnas.1108174108,cao2018unconventional}.

Effects similar to those of moir\'e lattices can also be expected for periodically deformed systems. Here, a periodic deformation can enlarge the unit cell. Periodic deformations, therefore, are expected to provide an alternative path to exciting new physical effects. A simple choice of periodic deformation field is given by
\begin{equation}
    \vec{U}_{base}(\vec r) =
    \begin{pmatrix}
        C \sin(\vec k_1 \cdot  \vec r) \\
        0 \\
    \end{pmatrix}
\end{equation}

\begin{figure}[H]
    \centering
\includegraphics[width=0.8\linewidth]{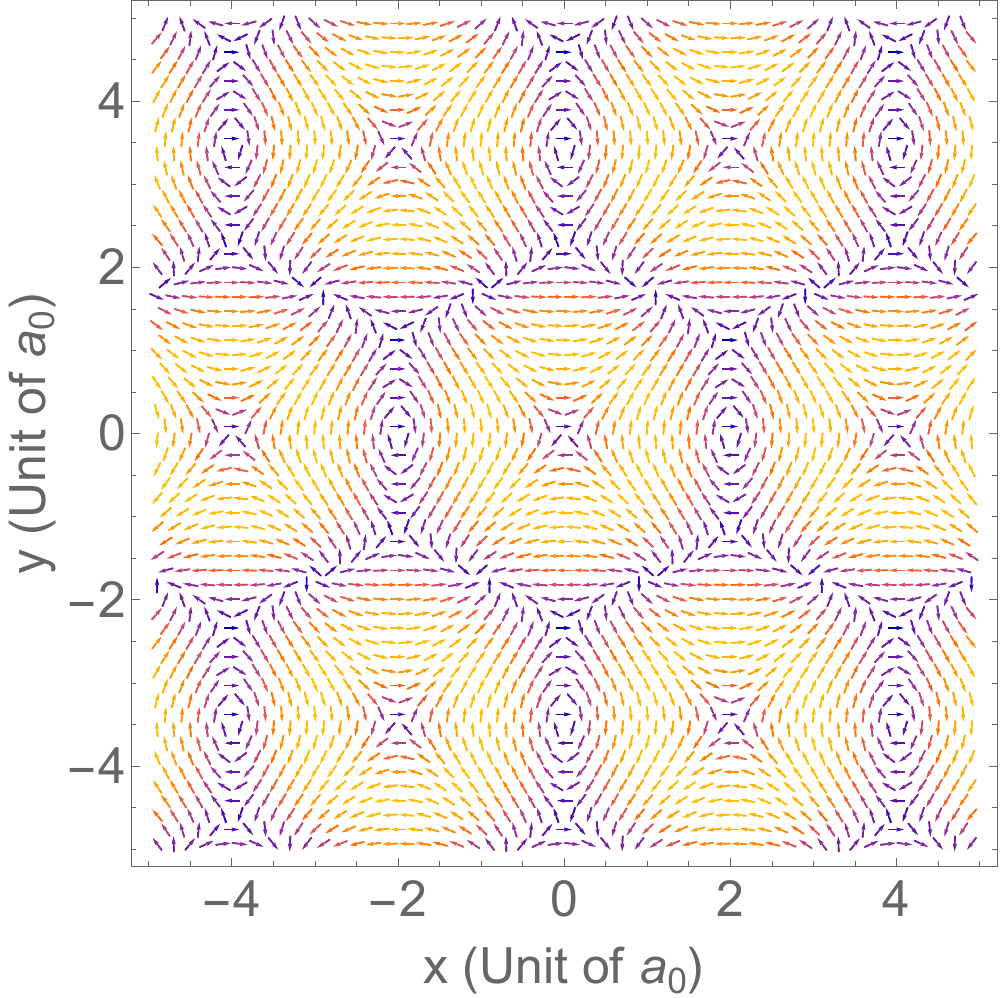}
    \caption{Hexagonally symmetric vector field that describes the deformation we consider in this work..} 
    \label{fig:deformation}
    \end{figure}

The undeformed Kagome lattice has a sixfold rotational symmetry. A deformation field that would allow an especially simple treatment would have the same symmetry. To build such a field, we can take a superposition of six terms rotated by $60^\circ$ as given by
\begin{equation}
    \vec{U}(\vec r) = \sum_{n = 0}^5R\left(\frac{n\pi}{3} \right) \vec U_{base}\left[R\left(\frac{n\pi}{3} \right)  \vec r\right].
\end{equation}
Fig. \ref{fig:deformation} shows a plot of our deformation field. 
It must be stressed that the sixfold symmetry of the deformation field should be understood as a symmetry of a vector field. Simply rotating the image in Fig.~\ref{fig:deformation} rotates the spatial positions while leaving the displayed displacement vectors unchanged, and therefore does not represent the relevant symmetry transformation. Instead, both the position and the displacement vector must be rotated. Explicitly, for a $60^\circ$ rotation $R=R(\pi/3)$,
\begin{equation}
    \vec U(R\vec r)=R\,\vec U(\vec r).
\end{equation}
Thus, Fig.~\ref{fig:deformation} is invariant under a $60^\circ$ rotation when both the spatial positions and the associated displacement vectors are rotated together.
Lastly we need to be fully concrete with our choice of deformation. Here and in all that follows we chose $\vec k_1 = (0, \frac{2\pi }{p \sqrt{3}})$ because it ensures a periodic lattice of sufficiently small size.

\begin{figure}[H]
    \centering
\includegraphics[width=\linewidth]{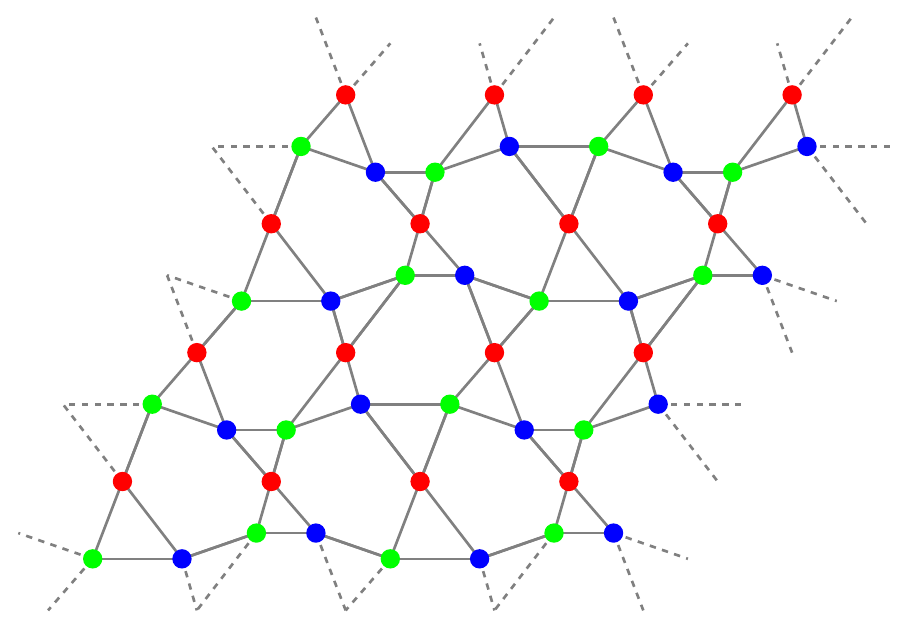}
    \caption{Deformed Kagome Lattice for deformation strength $C = 0.1$.} 
    \label{fig:Kagome:deformed}
    \end{figure}

We may now apply the deformation field to the Kagome lattice by computing new lattice locations $\vec{r}^\prime=\vec{r} +U (\vec r)$. In what follows, we choose relatively small distortion strengths such that nearest neighbors (NN) relations remain. The resulting lattice has a form like shown in Fig. \ref{fig:Kagome:deformed}.

\subsection{Deformed couplings and magnetic unit cell}

As the lattice gets deformed, the Heisenberg and DM interaction strengths are expected to change. That is, one would expect couplings to increase as sites get closer and weaker as they get further apart. A first-order approximation can be found via a Taylor series similar to \cite{Nayga_2022}. If we additionally assume rotational invariance, we obtain couplings that change as
    \begin{equation}
    \begin{aligned}
 J_{ij}&=J_0 \left[1-\beta_1\left(\frac{|a_{ij}|}{a_0}-1\right)\right]\\
 D_{ij}&=D_0 \left[1-\beta_2\left(\frac{|a_{ij}|}{a_0}-1\right)\right]
 \end{aligned}
 \label{eq:Deformed J}
    \end{equation}

Here, $J_0$ and $D_0$ are the coupling strengths of the Heisenberg and DM interactions in a non-deformed lattice. How strongly couplings react to the deformations is quantified by $\beta_i$. Moreover, $a_0$ is the non-deformed distance between neighboring sites, and $a_{ij}=|\vec r_i^\prime-\vec r_j^\prime|$ is the deformed distance between sites $i$ and $j$ (recall that $\vec r^\prime=\vec r+\vec U(\vec r)$).  We retain only bonds that are nearest neighbors in the undeformed Kagome lattice. The deformation strengths considered here are sufficiently small that this nearest-neighbor connectivity remains unchanged. Further-neighbor exchange interactions can be relevant in material-specific realizations of Kagome magnets \cite{Xie2021FeSn}, but are neglected in the present minimal model.

With these changes to the interaction strengths, spins on each site will naturally rearrange into a different ground-state structure. As in the undeformed case, to find the new ground state, we first computed the classical ground-state configuration for a finite $N_x \times N_y$ lattice with $N_x = N_y = 20$. We were then able to identify the unit cell of the new spin ground state. Our result (once we fixed the unit cell) is shown below in Fig. \ref{fig:Kagome_deformed_unitcell_spinstructure}.

\begin{figure}[H]
    \centering
    \includegraphics[width=\linewidth]{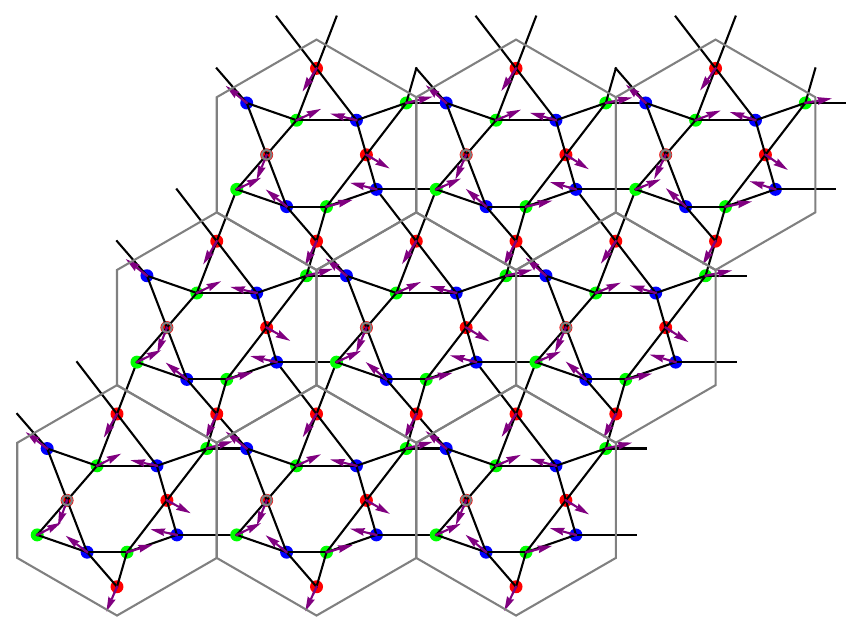}
    \caption{Deformed Kagome lattice for the antiferromagnetic exchange coupling $J = 1$ with deformation strength $C = 0.1$, parameters $\beta_1 = 2, \beta_2 = 2 $ , and DM interaction $|\frac{D_z}{J}| = 0.1$. Shown with purple arrows is the classical ground state configuration. We have also identified unit cells with a hexagonal tiling.}
    \label{fig:Kagome_deformed_unitcell_spinstructure}
    \end{figure}
Now it is important to distinguish the structural unit cell of the deformed lattice from the magnetic unit cell, defined as the smallest unit cell over which the classical spin configuration repeats. In principle, magnetic ordering could further enlarge the unit cell. Here, however, we find that the magnetic unit cell coincides with the enlarged 12-site structural unit cell generated by the deformation.

\subsection{Solution and bandstructure}
After we have found the classical ground state configuration of the deformed Kagome lattice, we follow the same procedure as in the undeformed case. That is, we first write each spin in a local frame where it points along the positive z-axis. Then, we apply the Holstein–Primakoff transformation and $1/S$ expansion to obtain a linear spin-wave theory in terms of bosons. The main change in our analysis is that the enlarged unit cell with 12 sites, rather than the original 3, leads to a larger Hamiltonian. Then, we may write the resulting Hamiltonian in momentum space after applying a Fourier transformation     
    \begin{align}
H = \sum_{\mathbf{k}}
\boldsymbol{\Psi}^\dag(\vec k)h(\vec k)
\boldsymbol{\Psi}(\vec k).
\end{align}
As usual, the effective single-body Bloch Hamiltonian can be read off as
\begin{align}
  h(\vec k) =  \begin{pmatrix}
A(\mathbf{k}) & B(\mathbf{k}) \\
B^{*}(-\mathbf{k}) & A^{*}(-\mathbf{k}) \notag
\end{pmatrix} 
\end{align}
Here, we made use of matrices $A(\vec k)$ and $B(\vec k)$, which are given as
\begin{equation}
\begin{aligned}
    A(\mathbf{k}) &= \frac{1}{4}\sum_{\langle i , j\rangle}\Gamma_{ij}^{(+)}e^{i \mathbf{k}\cdot \mathbf{r}_{ij}}\ket{i}\otimes\bra{j}+\frac{1}{2}\sum_{i}\Gamma_{ii}^{(z)}\ket{i}\otimes\bra{i}\\
     B(\mathbf{k}) &= \frac{1}{4}\sum_{\langle i , j\rangle}\Gamma_{ij}^{(-)}e^{i \mathbf{k}\cdot \mathbf{r}_{ij}}\ket{i}\otimes\bra{j}.
    \end{aligned}
\end{equation}
For the more general ground-state configurations encountered in the deformed system, the mixed components $\Gamma_{ij}^{xy}$ and $\Gamma_{ij}^{yx}$ need not vanish. We therefore use the generalized combinations
\begin{equation}
\begin{aligned}
&\Gamma_{ij}^{(\pm)}= \Gamma_{ij}^{xx}\pm\Gamma_{ij}^{yy}+i(\Gamma_{ij}^{yx}\mp\Gamma_{ij}^{xy})\quad  \Gamma_{ii}^{(z)} = -\sum_{j}N_{ij}\Gamma_{ij}^{zz}
\end{aligned}
\end{equation}
where in the first expression $i\neq j$. Moreover, we used the term $N_{ij}$, which equals $1$ if sites $i$ and $j$ are nearest neighbors (in the original undeformed lattice) and $0$ otherwise. We stress that $i$ and $j$ are labels for sites in the enlarged unit cell. Note also that the imaginary parts in $\Gamma_{ij}^{(\pm)}$ is due to treating ferromagnetic and antiferromagnetic cases simultaneously. We also stress that care has to be taken with $\vec r_{ij}=-\vec r_{ji}$, $\Gamma_{ji}^{xy}=\Gamma_{ij}^{yx}$ and $\Gamma_{j i}^{y x}=\Gamma_{i j}^{x y}$.

\noindent Here,  components are explicitly given as: 
\begin{equation}
\begin{aligned}
\Gamma_{ij}^{xx} &=   \cos\theta_i\cos\theta_j\Bigl[\cos(\phi_i-\phi_j)J_{ij} - D_{ij}^z\sin(\phi_i-\phi_j)\Bigr] \\ &+J_{ij}\sin\theta_i\sin\theta_j \\
\Gamma_{ij}^{xy} &= \cos\theta_i\Bigl[\cos(\phi_i-\phi_j)D_{ij}^z + J_{ij}\sin(\phi_i-\phi_j)\Bigr]\\
\Gamma_{ij}^{yx} &= -\cos\theta_j\Bigl[\cos(\phi_i-\phi_j)D_{ij}^z + J_{ij}\sin(\phi_i-\phi_j)\Bigr]\\
\Gamma_{ij}^{yy} &= \cos(\phi_i-\phi_j)J_{ij} - D_{ij}^z\sin(\phi_i-\phi_j)\\
\Gamma_{ij}^{zz} & =  \cos\theta_i\cos\theta_j\,J_{ij} \\ &+\sin\theta_i\Bigl[\cos(\phi_i-\phi_j)J_{ij} - D_{ij}^z\sin(\phi_i-\phi_j)\Bigr]\sin\theta_j
\end{aligned}
\end{equation}
Here, $\theta_i$ corresponds to the polar angle of the ground state spin $i$ and $\phi_i$ to the azimuthal angle.

\begin{figure}[H]
    \centering
    \subfloat[]{
\includegraphics[width=\linewidth]{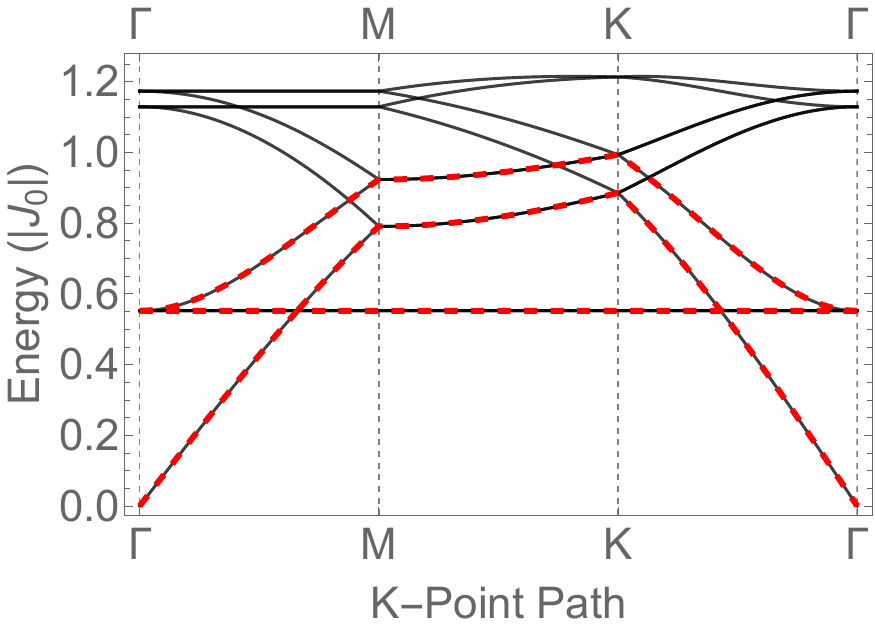}
    }
    \hfill
    \subfloat[]{
\includegraphics[width=\linewidth]{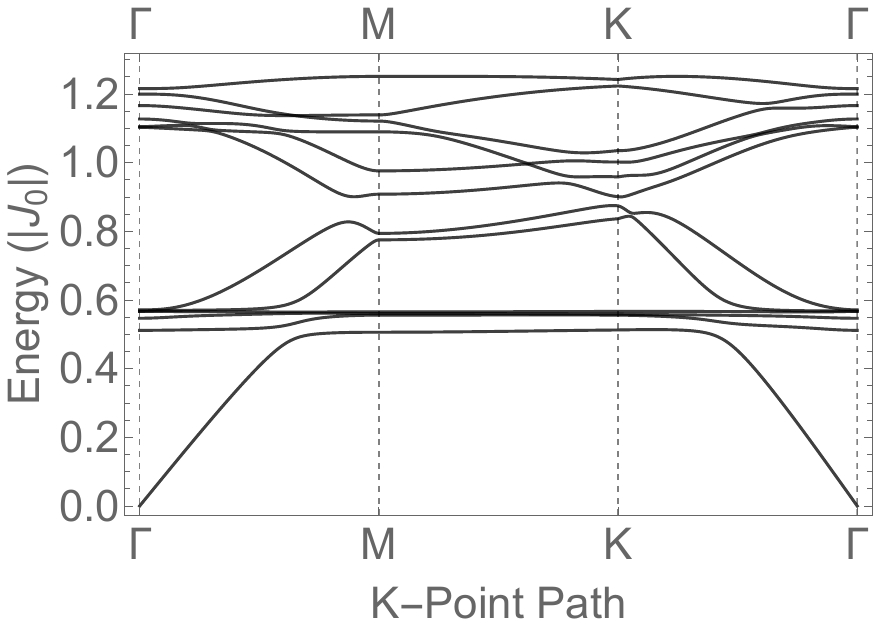}
    }
    \caption{Plot (a) shows the band structure for the antiferromagnetic undeformed case, where black lines are bands for the enlarged unit cells and red dashed lines are bands for the original unit cell. Plot (b) shows a band structure of the deformed case where distortion strength $C = 0.01$, $|\frac{D_z}{J_0}| = 0.1$, $\beta_1 = 7$, and $\beta_2 = 5$.}
\label{fig:band_undeformed_Kagome8}
\end{figure}

This Hamiltonian can again be diagonalized using paraunitary diagonalization. To get a better initial understanding for the anti-ferromagnetic case in Fig. \ref{fig:band_undeformed_Kagome8}(a) , we plotted the deformed case but with zero deformation strength and our undeformed case with the small unit cell.

We find that the red-dashed curve matches exactly some of the black bands. The reason for this effect is that the deformed case has an enlarged unit cell, leading to multiple copies of the same bands, just shifted in $k$-space. One set of band copies coincides exactly with the undeformed case from the smaller unit cell. This effect is the same as is known from the empty lattice approximation\cite{ashcroft1976solid}. It is important to distinguish this purely geometric band folding from the effects of the physical deformation. At zero deformation, the enlarged 12-site unit cell simply folds each of the three original Kagome bands into four branches in the reduced Brillouin zone, as shown in Fig.~\ref{fig:band_undeformed_Kagome8}(a). These folded branches exhibit crossings and no new deformation-induced gaps. The Chern numbers discussed below are instead calculated at finite deformation, where the modified couplings hybridize the folded branches and open gaps between them. The resulting isolated minibands are therefore genuine bands of the deformed Hamiltonian and can possess their own well-defined Chern numbers.

Next, we studied the effect of non-zero deformation on the band structure in Fig. \ref{fig:band_undeformed_Kagome8}(b). 
\begin{figure}[H]
    \centering
    \subfloat[]{
\includegraphics[width=\linewidth]{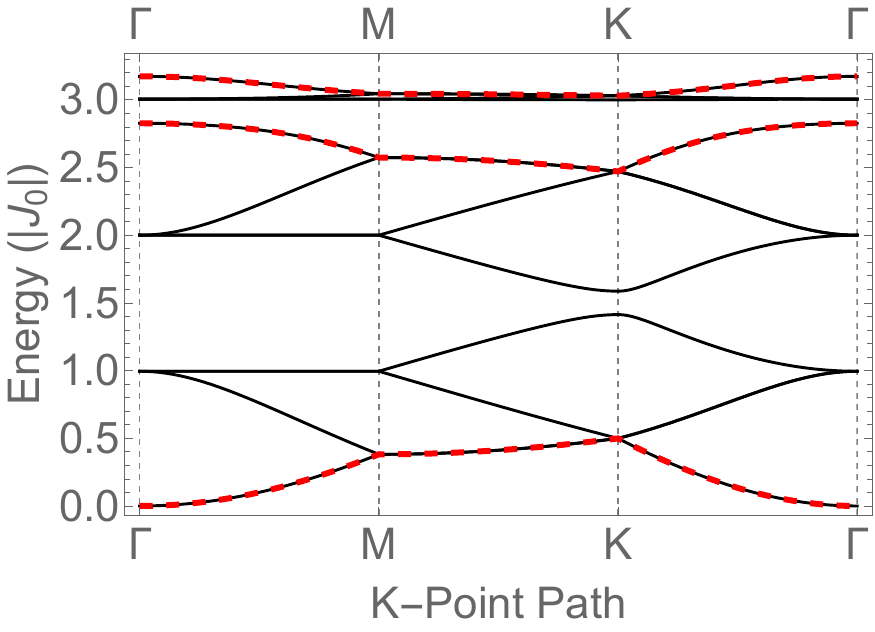}
    }
    \hfill
    \subfloat[]{
\includegraphics[width=\linewidth]{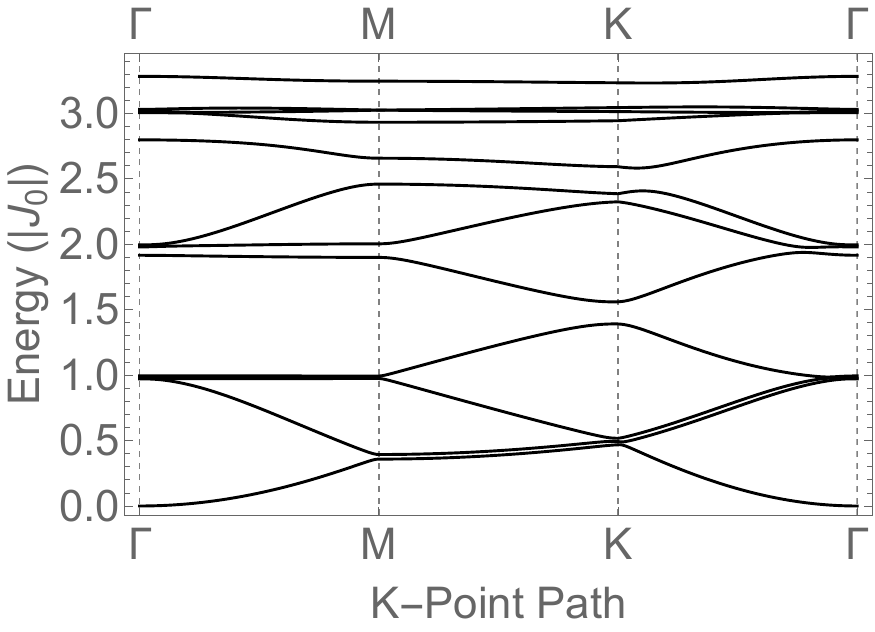}
    }
    \caption{Plot (a) shows the band structure for the ferromagnetic undeformed case, where black lines are bands for the enlarged unit cells and red dashed lines are for the original unit cell. Plot (b) shows a band structure of the deformed case where distortion strength $C = 0.01$, $|\frac{D_z}{J_0}| = 0.1$, $\beta_1 = 7$, and $\beta_2 = 5$.}
\label{fig:band_undeformed_Kagome}
\end{figure}  

Here, we observe multiple band gaps opening at the original crossings at the edges of the Brillouin zone. This kind of behavior is typical for weak distortions, as typically discussed in the context of the nearly free electron approximation that builds on the empty lattice approximation \cite{ashcroft1976solid}. While isolated bands make it tempting to search for Chern numbers, as we will see later, in the antiferromagnetic case, we regrettably obtain trivial results.

Given this idea, we transition to the ferromagnetic case. The analysis here parallels the antiferromagnetic scenario but is somewhat simpler, thanks to a collinear ground state with spins along the z-direction. Figure \ref{fig:band_undeformed_Kagome} shows both the undeformed and deformed band structures for comparison.

Again, in the upper plot, we observe perfect agreement between the zero-deformation case with the enlarged unit cell and the fully undeformed case. Additional bands are again due to an implicit empty lattice approximation. For the deformed bands in \ref{fig:band_undeformed_Kagome}(b), we again observe several band gaps that open.
Here, we choose the relatively weak deformation $C=0.01$ simply as an illustrative example to show the opening of gaps in the folded ferromagnetic band structure. Different parameter cuts are used below when studying the topological phase diagram. These band gaps, as we will see later, are crucial and lead to an interesting and varied topological structure.

\section{Band-topology}
\label{sec:IVbandtopo}
In addition to band structures alone, topological features of bands are also of much interest. Specifically, in our case, we are interested in Berry curvature and Chern numbers. Here, first it is useful to introduce a quantity called the Berry connection of band $n$\cite{Roberts_2024}

\begin{equation}  \mathbf{\mathcal{A}}_n(\mathbf{k})=-i\eta_{nn}\bra{u_{\mathbf{k}}}\eta \nabla_{\mathbf{k}}\ket{u_{\mathbf{k}}};\quad \eta=\mathrm{diag}(\mathbb{1},-\mathbb{1}),
\end{equation}
where vectors following the para unitary diagonalization procedure had to be "normalized" according to
\begin{equation}
    \bra{{u^m_{\mathbf{k}}}}\eta \ket{{u^n_{\mathbf{k}}}}=\eta_{mn}.
\end{equation}
Here, $u_{\mathbf{k}}$ are the Bloch eigenstates obtained from the paraunitary diagonalization procedure. The Berry curvature then is defined as 
\begin{equation}
    \mathbf{\Omega}(\mathbf{k}) = \nabla_{\mathbf{k}} \times \mathcal{A}(\mathbf{k}).
\end{equation}

When going around a closed loop $\mathcal{C}$ for a surface $S$ the Berry phase is 
\begin{equation}
      \gamma(\mathcal{C}) = \iint_S \mathbf{\Omega}(\mathbf{k})\cdot d\mathbf{S}
   \end{equation}

The Berry phase is a flux with Berry curvature as flux density. The aforementioned curve $\mathcal{C}$ for our purposes is the boundary of our Brillouin zone, and $S$ is the Brillouin zone itself. The so-called Chern number for the $n$th band is then defined as
\begin{equation}
    C_n = \frac{1}{2\pi}\iint_{BZ}\mathbf{\Omega}_n(\mathbf{k})\cdot d^2\mathbf{k},
\end{equation}
where the factor $1/(2\pi)$ is included to obtain integers\cite{thouless1982quantized}. 
Numerically, we evaluate the Chern numbers using the Fukui lattice-gauge algorithm \cite{Fukui2005}, generalized to the bosonic Bogoliubov problem by evaluating neighboring-state overlaps with the paraunitary metric $\eta$. Specifically, we worked on an $N_k\times N_k$ momentum-space grid, with $N_k=100$ used for the results presented in the manuscript. We verified convergence by increasing the grid density from $N_k=10$ to $N_k=150$. Away from band-touching points, the resulting Chern numbers converge to integers within numerical tolerance$= 10^{-10}$. At gap-closing points, the Berry curvature becomes singular and an individual-band Chern number is not well defined; we therefore identify these points from the closing of the corresponding direct band gap rather than assigning a Chern number at the transition itself. For an isolated spectral gap, one may also characterize the topology by the sum of the Chern numbers of the bands on one side of the gap. Such cumulative Chern numbers are particularly useful in connection with bulk-boundary correspondence and the net chirality of edge modes crossing a gap \cite{PhysRevB.87.174427}. In the present bosonic Bogoliubov problem, however, the lowest positive-energy band touches its negative-norm partner at zero energy at the $\Gamma$ point and therefore does not possess an individually well-defined Chern number. Consequently, a straightforward cumulative sum starting from the lowest positive-energy band cannot be constructed from the individual band Chern numbers considered here, and we do not pursue gap Chern numbers further in this work.

The individual band topology nevertheless remains physically meaningful. In particular, each isolated magnon band carries its own Berry curvature and Chern number, and the band-resolved Berry curvature enters bosonic transport coefficients with an energy- and temperature-dependent Bose occupation factor \cite{Roberts_2024,MatsumotoMurakami2011,laurell2018magnon}. From a semiclassical perspective, the Berry curvature also enters the anomalous velocity \cite{RevModPhys.82.1959}, which is relevant for transverse transport. Thus, unlike in a fermionic system with completely filled bands below a Fermi energy, the transport response is not determined solely by a cumulative Chern number.

    \subsection{Antiferromagnetic case}
First, we studied the antiferromagnetic case $(J = 1)$ and observed that, in all tested cases, the Chern numbers of all bands are trivial. Importantly, one can prove that the result holds in general. Specifically, the Hamiltonian is invariant under a combined symmetry transformation of a $180^\circ$ degree rotation around the $z$-axis and time inversion symmetry. Importantly, this symmetry survives in low-energy spin-wave theory, since the ground-state spin configuration is also invariant under it. The net effect is that the Berry curvature is antisymmetric, so the integral over the Brillouin zone vanishes. A detailed proof is given in appendix \ref{app:berry_curv-antisymm_for_antiferr}

This result made the antiferromagnetic case uninteresting from a topological perspective. 
 \subsection{Ferromagnetic case}
The ferromagnetic case $(J = -1)$ has a classical ground state configuration that is along the z-direction and therefore spontaneously breaks the symmetry that for the antiferromagnetic case led to zero Chern number. We can thus hope for non-trivial Chern numbers. As a first check, we calculated the Chern numbers for the undeformed case and found $-1$, $0$, $1$, which agree with well-known literature results \cite{mook2014edge}.  

Moving to the deformed case, we again find that the magnetic unit cell coincides with the enlarged 12-site structural unit cell of the deformed lattice, as in the antiferromagnetic case. 

\begin{figure}[h!]
    \centering
    \subfloat[]{
\includegraphics[width=0.9\linewidth]{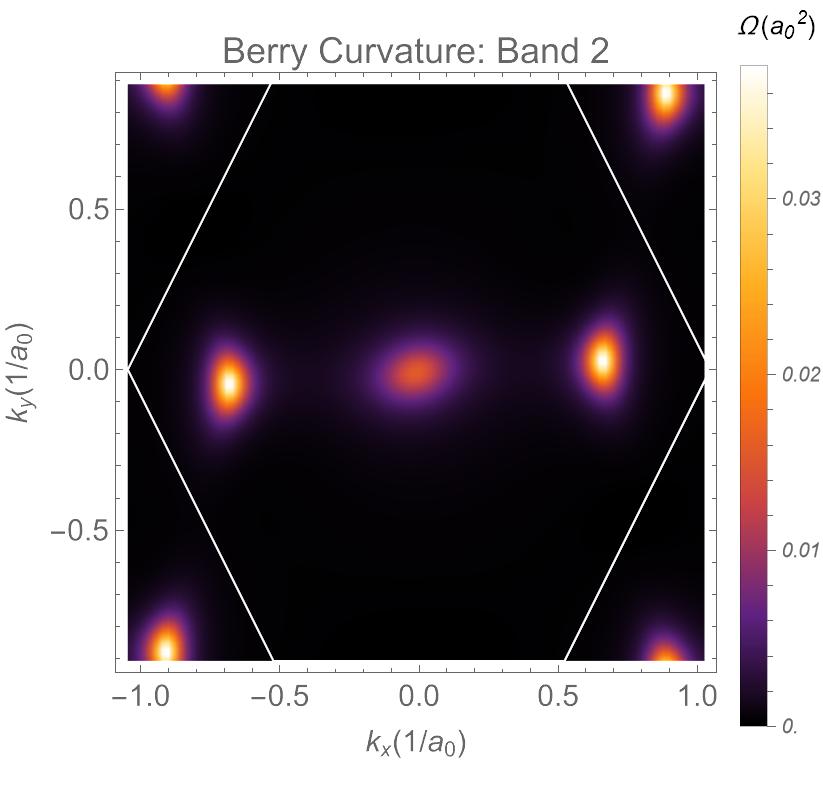}
    }
    \hfill
    \subfloat[]{
\includegraphics[width=0.9\linewidth]{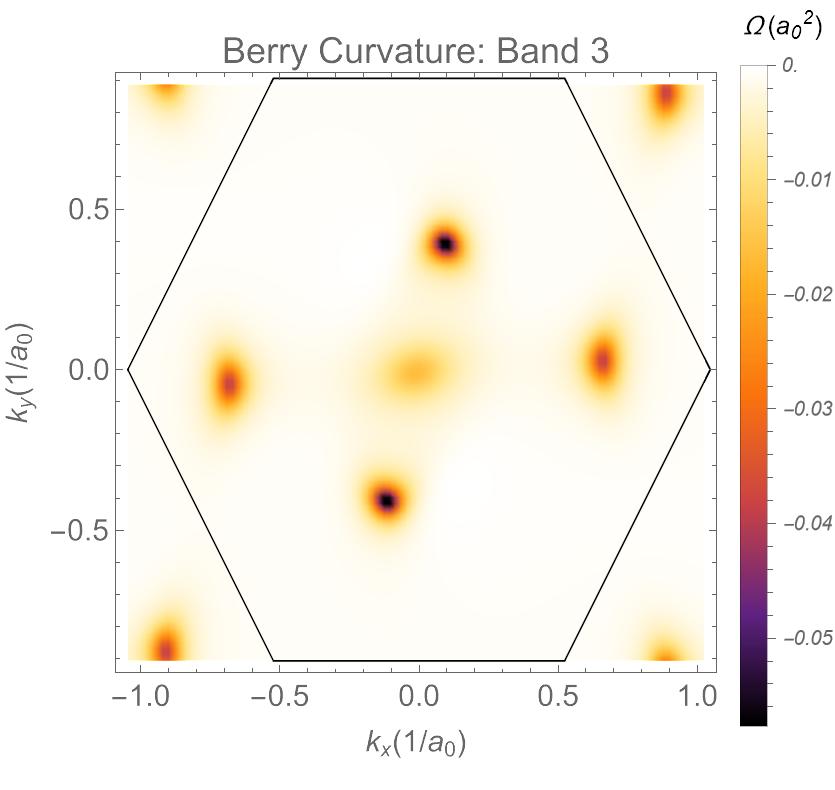}
    }
    \caption{The first plot on the left shows the berry curvature of band 2, and the second one for band 3. The chosen Hamiltonian parameters are $D/J = -0.2$, $\beta_1 = 7$ and $\beta_2 = 5$.}
    \label{fig:Berry curvature}
\end{figure}

As a first  step in our investigation, we then studied the Berry curvature of different bands, as shown in Fig. \ref{fig:Berry curvature} for bands 2 and 3.

We observe that the Berry curvature of band 2 mostly has positive contributions, which are dominated by two bright spots in the Brillouin zone. The associated Chern  for this case is 1. In the lower plot for band 3, we see 4 dark spots with a large negative contribution, and the band has Chern number -2. Already here, we see that the deformed system supports a broader range of Chern numbers than the undeformed case.

\begin{figure*}[t]
    \centering
\includegraphics[width=0.95\textwidth]{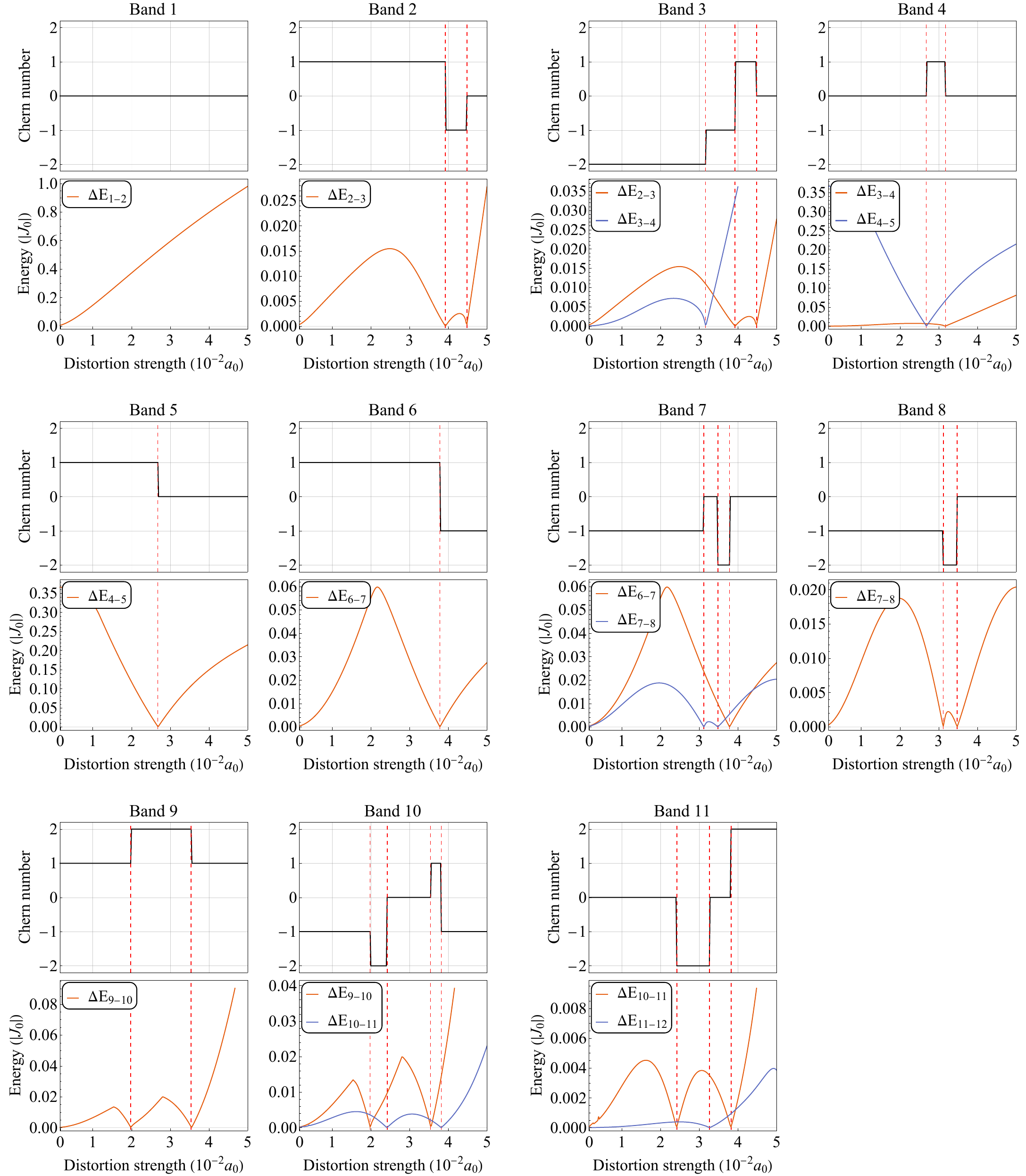}
 \captionsetup{
        width=0.95\textwidth,
        justification=centering,
        singlelinecheck=false,
        margin=2pt
    }
    \caption{Plots of Chern number and band gap for various bands as a function of distortion strength $C$. The Hamiltonian has other parameters fixed as $\frac{D_Z}{J} = -0.2$, $\beta_1 = 7$, and $\beta_2 = 5$.}
    \label{fig: vs Distortion strength/plot}
\end{figure*}

Motivated by this interesting result, we next examine the impact of deformation strength $C$ on Chern numbers. We set fixed parameters to $D_z/J = -0.2$, $\beta_1 = 7$, and $\beta_2 = 5$ in the Hamiltonian. These values are motivated by the fact that the DM interaction $D_z$ is typically weaker than the exchange coupling $J$. Moreover, we have chosen values for $\beta_i$ that lie in the parameter range that was motivated in \cite{Nayga_2022}.  In our plots of the Chern numbers, we include all positive-energy bands except band 12. Although the negative-energy branches are not additional physical magnon excitations, band 12 touches its negative-norm Bogoliubov partner at zero energy at the $\Gamma$ point. Band 12 is therefore not spectrally isolated, and an individual Chern number cannot be assigned to it in the usual manner. This zero-energy degeneracy does not affect the Chern numbers of the remaining positive-energy bands as long as they remain spectrally isolated. At parameter values where such bands touch one another, their individual Chern numbers are correspondingly ill-defined and these points identify the topological transition boundaries. Results with accompanying plots for the band gaps are shown in Fig. \ref{fig: vs Distortion strength/plot}.

We immediately observe a plethora of topological phase transitions, each accompanied by a band-gap closing. Band 1 remains topologically trivial throughout, while bands 2--11 undergo multiple transitions as they touch adjacent bands. In particular, band 3 reaches Chern number $-2$, band 9 reaches $+2$, and bands 7, 8, 10, and 11 also access phases with $|C_n|=2$. Several bands pass between trivial and nontrivial phases, while others undergo direct transitions between different nonzero Chern numbers. Overall, the deformation redistributes Berry curvature among neighboring bands and produces repeated topological transitions across the spectrum. Higher-Chern-number magnon phases have also been reported for Kagome ferromagnets with additional anisotropic interactions \cite{Ni2025}. Here, they arise instead within a minimal Heisenberg--DM model through the two-dimensional periodic deformation.

\begin{figure*}[t]
    \centering
\includegraphics[width=0.95\textwidth]{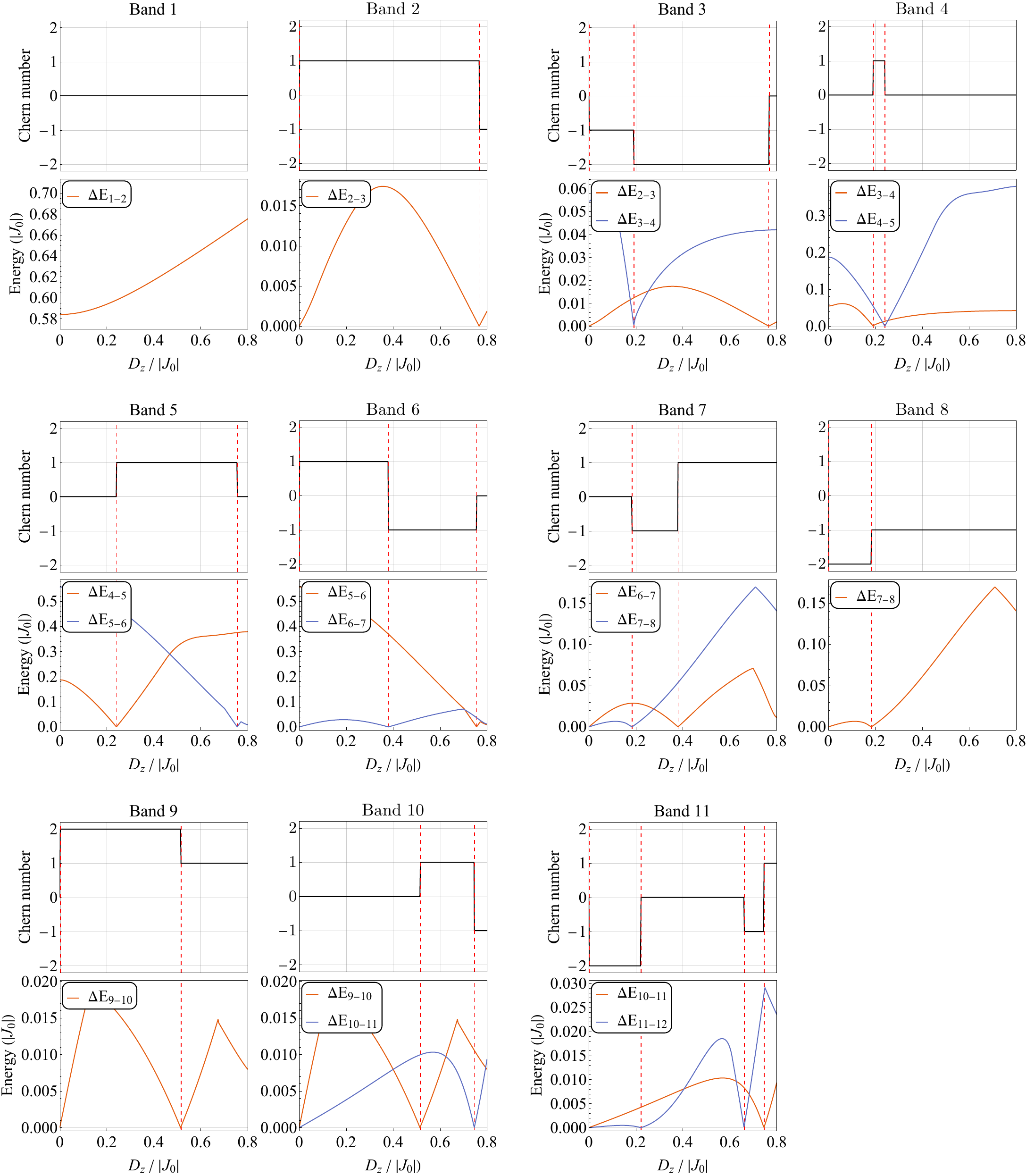}
 \captionsetup{
        width=0.95\textwidth,
        justification=centering,
        singlelinecheck=false,
        margin=2pt
    }
    \caption{Plots of Chern number and band gap for various bands as a function of DM interaction strength $D_z/J$. The Hamiltonian has other parameters fixed as $C=0.03$, $\beta_1 = 7$, and $\beta_2 = 5$.}
    \label{fig:vs Distortion DMI/plot}
\end{figure*}

After observing that deformations trigger a plethora of topological phase transitions, we also want to see whether they make the system more sensitive to changes in other parameters. That is, we study phase transitions and band-gap closings that occur as we vary the strength of the DM interaction. For this parameter cut, we fix the deformation strength to $C=0.03$ and set $\beta_1=7$ and $\beta_2=5$. This choice is distinct from the illustrative band structure in Fig.~\ref{fig:band_undeformed_Kagome}(b), where $C=0.01$. The resulting plots of Chern numbers and band gaps are shown in Fig. \ref{fig:vs Distortion DMI/plot}.

As in the deformation sweep, we observe numerous band touchings accompanied by changes in Chern number. Band 1 remains trivial, whereas bands 2--11 undergo several transitions as the DM interaction is varied. Notably, bands 3, 8, 9, and 11 access phases with $|C_n|=2$, while several other bands alternate between trivial and nontrivial phases. The pattern differs quantitatively from that obtained by varying the deformation strength, showing that the periodically deformed system supports a broad range of topological phases as the DM interaction is tuned.

In summary, we have observed that the topological phase diagram is enriched by the deformation. Suggesting that already topological magnetic systems can display an even more varied topological structure.

\section{Conclusion}
\label{sec:conclusion}

In conclusion, we have investigated a  minimal nearest-neighbor Heisenberg model with Dzyaloshinskii–Moriya (DM) interaction on a periodically deformed Kagome lattice. The deformation, implemented via a periodic hexagonal vector field, enlarges the unit cell and consequently enriches the band structure, leading to multiple band-gap openings in the spectrum. Our analysis of the band topology reveals that, in the antiferromagnetic case $J = 1$, the system exhibits a trivial topology, protected by an underlying symmetry that remains unbroken in the ground state. In contrast, for the ferromagnetic case $J = -1$, this symmetry is spontaneously broken by the classical ground state, giving rise to a complex topological structure characterized by multiple phase transitions. The nontrivial magnon-band topology identified here may have consequences for transverse magnon transport, such as the thermal Hall response; however, a quantitative determination of such effects requires a separate calculation of the corresponding transport coefficients and is beyond the scope of the present work. A natural extension of the present minimal model would be to include further-neighbor exchange interactions, which may quantitatively shift the phase boundaries and generate additional topological features in material-specific realizations. Looking ahead, an important direction for future work is the experimental realization of such lattice deformations, with cold-atom platforms representing promising candidates. It would also be very interesting to see what effects one may achieve by other kind of deformations. For instance, can a uniaxial strain lead to something similar to Landau levels? In addition, one could add electromagnetic fields, such as circularly polarized light, and observe their combined effects with the deformations.

In summary, our results demonstrate, with a relatively simple example, that deformed magnetic systems can host a rich variety of topological phenomena, particularly when the corresponding undeformed system already exhibits nontrivial topology.
 \section{Acknowledgments}
 M.V. gratefully acknowledges the support provided by the Deanship of Research Oversight and Coordination (DROC) and the Interdisciplinery Reasearch Center(IRC) for Advanced Quantum Computing (AQC) at King Fahd University of Petroleum \& Minerals (KFUPM) for funding their contribution to this work through research grant No. INQC2607.

\bibliographystyle{unsrt}
\bibliography{literature}

@article{laurell2018magnon,
  title={Magnon thermal Hall effect in kagome antiferromagnets with Dzyaloshinskii-Moriya interactions},
  author={Laurell, Pontus and Fiete, Gregory A},
  journal={Physical Review B},
  volume={98},
  number={9},
  pages={094419},
  year={2018},
  publisher={APS}
}

@article{Roberts_2024,
   title={Geometric origin of the intrinsic transverse spin transport in a canted-antiferromagnet/heavy-metal heterostructure},
   volume={109},
   ISSN={2469-9969},
   url={http://dx.doi.org/10.1103/PhysRevB.109.174436},
   DOI={10.1103/physrevb.109.174436},
   number={17},
   journal={Physical Review B},
   publisher={American Physical Society (APS)},
   author={Roberts, Wesley and Ma, Bowen and Rodriguez-Vega, Martin and Fiete, Gregory A.},
   year={2024},
   month=may }

@article{moriya1960anisotropic,
  title={Anisotropic superexchange interaction and weak ferromagnetism},
  author={Moriya, T{\^o}ru},
  journal={Physical review},
  volume={120},
  number={1},
  pages={91},
  year={1960},
  publisher={APS}
}

@article{wen1990topological,
  title={Topological orders in rigid states},
  author={Wen, Xiao-Gang},
  journal={International Journal of Modern Physics B},
  volume={4},
  number={02},
  pages={239--271},
  year={1990},
  publisher={World Scientific}
}

@article{thouless1982quantized,
  title={Quantized Hall conductance in a two-dimensional periodic potential},
  author={Thouless, David J and Kohmoto, Mahito and Nightingale, M Peter and den Nijs, Marcel},
  journal={Physical review letters},
  volume={49},
  number={6},
  pages={405},
  year={1982},
  publisher={APS}
}

@article{fert2017magnetic,
  title={Magnetic skyrmions: advances in physics and potential applications},
  author={Fert, Albert and Reyren, Nicolas and Cros, Vincent},
  journal={Nature Reviews Materials},
  volume={2},
  number={7},
  pages={17031},
  year={2017},
  publisher={Nature Publishing Group}
}

@article{nayak2008non,
  title={Non-Abelian anyons and topological quantum computation},
  author={Nayak, Chetan and Simon, Steven H and Stern, Ady and Freedman, Michael and Das Sarma, Sankar},
  journal={Reviews of Modern Physics},
  volume={80},
  number={3},
  pages={1083--1159},
  year={2008},
  publisher={APS}
}

@article{Heisenberg1928,
  author  = {Heisenberg, Werner},
  title   = {Zur Theorie des Ferromagnetismus},
  journal = {Zeitschrift f{\"u}r Physik},
  volume  = {49},
  number  = {9},
  pages   = {619--636},
  year    = {1928},
  doi     = {10.1007/BF01328601}
}

@article{bloch1930theorie,
  title={Zur theorie des ferromagnetismus},
  author={Bloch, Felix},
  journal={Zeitschrift f{\"u}r Physik},
  volume={61},
  number={3},
  pages={206--219},
  year={1930},
  publisher={Springer}
}

@article{owerre2018strain,
  title={Strain-induced topological magnon phase transitions: Applications to kagome-lattice ferromagnets},
  author={Owerre, SA},
  journal={Journal of Physics: Condensed Matter},
  volume={30},
  number={24},
  pages={245803},
  year={2018},
  publisher={IOP Publishing}
}

@article{baibich1988giant,
  title={Giant magnetoresistance of (001) Fe/(001) Cr magnetic superlattices},
  author={Baibich, Mario Norberto and Broto, Jean Marc and Fert, Albert and Van Dau, F Nguyen and Petroff, Fr{\'e}d{\'e}ric and Etienne, P and Creuzet, G and Friederich, A and Chazelas, J},
  journal={Physical review letters},
  volume={61},
  number={21},
  pages={2472},
  year={1988},
  publisher={APS}
}

@article{binasch1989enhanced,
  title={Enhanced magnetoresistance in layered magnetic structures with antiferromagnetic interlayer exchange},
  author={Binasch, Gr{\"u}nberg and Gr{\"u}nberg, Peter and Saurenbach, F and Zinn, W},
  journal={Physical review B},
  volume={39},
  number={7},
  pages={4828},
  year={1989},
  publisher={APS}
}

@article{yuasa2007giant,
  title={Giant tunnel magnetoresistance in magnetic tunnel junctions with a crystalline MgO (0 0 1) barrier},
  author={Yuasa, S and Djayaprawira, DD},
  journal={Journal of Physics D: Applied Physics},
  volume={40},
  number={21},
  pages={R337--R354},
  year={2007}
}

@article{bhatti2017spintronics,
  title={Spintronics based random access memory: a review},
  author={Bhatti, Sabpreet and Sbiaa, Rachid and Hirohata, Atsufumi and Ohno, Hideo and Fukami, Shunsuke and Piramanayagam, SN},
  journal={Materials today},
  volume={20},
  number={9},
  pages={530--548},
  year={2017},
  publisher={Elsevier}
}

@article{nakatani2018read,
  title={Read sensor technology for ultrahigh density magnetic recording},
  author={Nakatani, Tomoya and Gao, Zheng and Hono, Kazuhiro},
  journal={MRS Bulletin},
  volume={43},
  number={2},
  pages={106--111},
  year={2018},
  publisher={Cambridge University Press}
}

@article{moodera1995large,
  title={Large magnetoresistance at room temperature in ferromagnetic thin film tunnel junctions},
  author={Moodera, Jagadeesh Subbaiah and Kinder, Lisa R and Wong, Terrilyn M and Meservey, R},
  journal={Physical review letters},
  volume={74},
  number={16},
  pages={3273},
  year={1995},
  publisher={APS}
}

@article{miyazaki1995giant,
  title={Giant magnetic tunneling effect in Fe/Al2O3/Fe junction},
  author={Miyazaki, Terunobu and Tezuka, Nobuki},
  journal={Journal of magnetism and magnetic materials},
  volume={139},
  number={3},
  pages={L231--L234},
  year={1995},
  publisher={Elsevier}
}

@article{yuasa2004giant,
  title={Giant room-temperature magnetoresistance in single-crystal Fe/MgO/Fe magnetic tunnel junctions},
  author={Yuasa, Shinji and Nagahama, Taro and Fukushima, Akio and Suzuki, Yoshishige and Ando, Koji},
  journal={Nature materials},
  volume={3},
  number={12},
  pages={868--871},
  year={2004},
  publisher={Nature Publishing Group UK London}
}

@article{parkin2004giant,
  title={Giant tunnelling magnetoresistance at room temperature with MgO (100) tunnel barriers},
  author={Parkin, Stuart SP and Kaiser, Christian and Panchula, Alex and Rice, Philip M and Hughes, Brian and Samant, Mahesh and Yang, See-Hun},
  journal={Nature materials},
  volume={3},
  number={12},
  pages={862--867},
  year={2004},
  publisher={Nature Publishing Group UK London}
}

@article{dieny1994giant,
  title={Giant magnetoresistance in spin-valve multilayers},
  author={Di{\'e}ny, Bernard},
  journal={Journal of Magnetism and Magnetic Materials},
  volume={136},
  number={3},
  pages={335--359},
  year={1994},
  publisher={Elsevier}
}

@article{savary2017quantum,
  title={Quantum spin liquids: a review},
  author={Savary, Lucile and Balents, Leon},
  journal={Reports on Progress in Physics},
  volume={80},
  number={1},
  pages={016502},
  year={2017},
  publisher={IOP Publishing}
}

@article{kitaev2006anyons,
  title={Anyons in an exactly solved model and beyond},
  author={Kitaev, Alexei},
  journal={Annals of Physics},
  volume={321},
  number={1},
  pages={2--111},
  year={2006},
  publisher={Elsevier}
}

@article{cao2018unconventional,
  title={Unconventional superconductivity in magic-angle graphene superlattices},
  author={Cao, Yuan and Fatemi, Valla and Fang, Shiang and Watanabe, Kenji and Taniguchi, Takashi and Kaxiras, Efthimios and Jarillo-Herrero, Pablo},
  journal={Nature},
  volume={556},
  number={7699},
  pages={43--50},
  year={2018},
  publisher={Nature Publishing Group UK London}
}

@article{guinea2010energy,
  title={Energy gaps and a zero-field quantum Hall effect in graphene by strain engineering},
  author={Guinea, Francisco and Katsnelson, Mikhail I and Geim, AK},
  journal={Nature Physics},
  volume={6},
  number={1},
  pages={30--33},
  year={2010},
  publisher={Nature Publishing Group UK London}
}

@article{mook2014edge,
  title={Edge states in topological magnon insulators},
  author={Mook, Alexander and Henk, J{\"u}rgen and Mertig, Ingrid},
  journal={Physical Review B},
  volume={90},
  number={2},
  pages={024412},
  year={2014},
  publisher={APS}
}

@article{landau1937theory,
  title={On the theory of phase transitions},
  author={Landau, Lev Davidovich and others},
  journal={Zh. eksp. teor. Fiz},
  volume={7},
  number={19-32},
  pages={926},
  year={1937}
}

@article{klitzing1980new,
  title={New method for high-accuracy determination of the fine-structure constant based on quantized Hall resistance},
  author={Klitzing, K v and Dorda, Gerhard and Pepper, Michael},
  journal={Physical review letters},
  volume={45},
  number={6},
  pages={494},
  year={1980},
  publisher={APS}
}

@article{dzyaloshinsky1958thermodynamic,
  title={A thermodynamic theory of “weak” ferromagnetism of antiferromagnetics},
  author={Dzyaloshinsky, Igor},
  journal={Journal of physics and chemistry of solids},
  volume={4},
  number={4},
  pages={241--255},
  year={1958},
  publisher={Elsevier}
}

@article{holstein1940field,
  title={Field dependence of the intrinsic domain magnetization of a ferromagnet},
  author={Holstein, Theodore and Primakoff, Henry},
  journal={Physical Review},
  volume={58},
  number={12},
  pages={1098},
  year={1940},
  publisher={APS}
}

@article{COLPA1978327,
title = {Diagonalization of the quadratic boson hamiltonian},
journal = {Physica A: Statistical Mechanics and its Applications},
volume = {93},
number = {3},
pages = {327-353},
year = {1978},
issn = {0378-4371},
doi = {https://doi.org/10.1016/0378-4371(78)90160-7},
url = {https://www.sciencedirect.com/science/article/pii/0378437178901607},
author = {J.H.P. Colpa}
}

@article{PhysRevB.87.174427,
  title = {Topological chiral magnonic edge mode in a magnonic crystal},
  author = {Shindou, Ryuichi and Matsumoto, Ryo and Murakami, Shuichi and Ohe, Jun-ichiro},
  journal = {Phys. Rev. B},
  volume = {87},
  issue = {17},
  pages = {174427},
  numpages = {11},
  year = {2013},
  month = {May},
  publisher = {American Physical Society},
  doi = {10.1103/PhysRevB.87.174427},
  url = {https://link.aps.org/doi/10.1103/PhysRevB.87.174427}
}

@article{
doi:10.1073/pnas.1108174108,
author = {Rafi Bistritzer  and Allan H. MacDonald },
title = {Moiré bands in twisted double-layer graphene},
journal = {Proceedings of the National Academy of Sciences},
volume = {108},
number = {30},
pages = {12233-12237},
year = {2011},
doi = {10.1073/pnas.1108174108},
URL = {https://www.pnas.org/doi/abs/10.1073/pnas.1108174108},
eprint = {https://www.pnas.org/doi/pdf/10.1073/pnas.1108174108}}

@article{miftah_saud_hbngraph,
  title = {Light-induced transitions of valley Chern numbers and flat bands in a nontwisted moir\'e graphene--hexagonal boron nitride superlattice},
  author = {Alabdulal, Saud and Anfa, Miftah Hadi Syahputra and Bahlouli, Hocine and Vogl, Michael},
  journal = {Phys. Rev. B},
  volume = {112},
  issue = {19},
  pages = {195411},
  numpages = {9},
  year = {2025},
  month = {Nov},
  publisher = {American Physical Society},
  doi = {10.1103/4zcv-2c1n},
  url = {https://link.aps.org/doi/10.1103/4zcv-2c1n}
}

@article{PhysRevB.90.155406,
  title = {Electronic properties of graphene/hexagonal-boron-nitride moir\'e superlattice},
  author = {Moon, Pilkyung and Koshino, Mikito},
  journal = {Phys. Rev. B},
  volume = {90},
  issue = {15},
  pages = {155406},
  numpages = {12},
  year = {2014},
  month = {Oct},
  publisher = {American Physical Society},
  doi = {10.1103/PhysRevB.90.155406},
  url = {https://link.aps.org/doi/10.1103/PhysRevB.90.155406}
}

@book{ashcroft1976solid,
  title={Solid state physics},
  author={Ashcroft, Neil W and Mermin, N David and others},
  year={1976},
  publisher={holt, rinehart and winston, new york London}
}

@article{RevModPhys.82.1959,
  title = {Berry phase effects on electronic properties},
  author = {Xiao, Di and Chang, Ming-Che and Niu, Qian},
  journal = {Rev. Mod. Phys.},
  volume = {82},
  issue = {3},
  pages = {1959--2007},
  numpages = {0},
  year = {2010},
  month = {Jul},
  publisher = {American Physical Society},
  doi = {10.1103/RevModPhys.82.1959},
  url = {https://link.aps.org/doi/10.1103/RevModPhys.82.1959}
}

@article{Nayga_2022,
   title={Strain tuning of highly frustrated magnets: Order and disorder in the distorted kagome Heisenberg antiferromagnet},
   volume={105},
   ISSN={2469-9969},
   url={http://dx.doi.org/10.1103/PhysRevB.105.094426},
   DOI={10.1103/physrevb.105.094426},
   number={9},
   journal={Physical Review B},
   publisher={American Physical Society (APS)},
   author={Nayga, Mary Madelynn and Vojta, Matthias},
   year={2022},
   month=mar }

@article{PhysRevResearch.2.043243,
  title = {Resummation of the Holstein-Primakoff expansion and differential equation approach to operator square roots},
  author = {Vogl, Michael and Laurell, Pontus and Zhang, Hao and Okamoto, Satoshi and Fiete, Gregory A.},
  journal = {Phys. Rev. Res.},
  volume = {2},
  issue = {4},
  pages = {043243},
  numpages = {11},
  year = {2020},
  month = {Nov},
  publisher = {American Physical Society},
  doi = {10.1103/PhysRevResearch.2.043243},
  url = {https://link.aps.org/doi/10.1103/PhysRevResearch.2.043243}
}

@Article{10.21468/SciPostPhys.10.1.007,
	title={{Newton series expansion of bosonic operator functions}},
	author={Jürgen König and Alfred Hucht},
	journal={SciPost Phys.},
	volume={10},
	pages={007},
	year={2021},
	publisher={SciPost},
	doi={10.21468/SciPostPhys.10.1.007},
	url={https://scipost.org/10.21468/SciPostPhys.10.1.007},
}

@article{c3n8-1h7f,
  title = {High-accuracy evaluation of nonthermal magnetic states beyond spin-wave theory: Applications to higher-energy states},
  author = {Roberts, Wesley and Vogl, Michael and Moessner, Roderich and Fiete, Gregory A.},
  journal = {Phys. Rev. B},
  volume = {112},
  issue = {21},
  pages = {214440},
  numpages = {14},
  year = {2025},
  month = {Dec},
  publisher = {American Physical Society},
  doi = {10.1103/c3n8-1h7f},
  url = {https://link.aps.org/doi/10.1103/c3n8-1h7f}
}

@article{Xie2021FeSn,
author={Xie, Yaofeng
and Chen, Lebing
and Chen, Tong
and Wang, Qi
and Yin, Qiangwei
and Stewart, J. Ross
and Stone, Matthew B.
and Daemen, Luke L.
and Feng, Erxi
and Cao, Huibo
and Lei, Hechang
and Yin, Zhiping
and MacDonald, Allan H.
and Dai, Pengcheng},
title={Spin excitations in metallic kagome lattice FeSn and CoSn},
journal={Communications Physics},
year={2021},
month={Nov},
day={11},
volume={4},
number={1},
pages={240},
issn={2399-3650},
doi={10.1038/s42005-021-00736-8},
url={https://doi.org/10.1038/s42005-021-00736-8}
}

@article{MatsumotoMurakami2011,
  title = {Rotational motion of magnons and the thermal Hall effect},
  author = {Matsumoto, Ryo and Murakami, Shuichi},
  journal = {Phys. Rev. B},
  volume = {84},
  issue = {18},
  pages = {184406},
  numpages = {9},
  year = {2011},
  month = {Nov},
  publisher = {American Physical Society},
  doi = {10.1103/PhysRevB.84.184406},
  url = {https://link.aps.org/doi/10.1103/PhysRevB.84.184406}
}

@article{Ni2025,
  title = {Multiple topological phases of magnons induced by Dzyaloshinskii-Moriya and pseudodipolar anisotropic exchange interactions in kagome ferromagnets},
  author = {Ni, Jin-Yu and Zheng, Xia-Ming and Wei, Peng-Tao and Liu, Da-Yong and Zou, Liang-Jian},
  journal = {Phys. Rev. B},
  volume = {112},
  issue = {18},
  pages = {184431},
  numpages = {17},
  year = {2025},
  month = {Nov},
  publisher = {American Physical Society},
  doi = {10.1103/gr5v-26np},
  url = {https://link.aps.org/doi/10.1103/gr5v-26np}
}

@article{Fukui2005,
author = {Fukui ,Takahiro and Hatsugai ,Yasuhiro and Suzuki ,Hiroshi},
title = {Chern Numbers in Discretized Brillouin Zone: Efficient Method of Computing (Spin) Hall Conductances},
journal = {Journal of the Physical Society of Japan},
volume = {74},
number = {6},
pages = {1674-1677},
year = {2005},
doi = {10.1143/JPSJ.74.1674},

URL = {   https://doi.org/10.1143/JPSJ.74.1674
},
eprint = {  https://doi.org/10.1143/JPSJ.74.1674 }
}
\appendix
\section{Anti-symmetric Berry curvature in the anti-ferromagnetic case}
\label{app:berry_curv-antisymm_for_antiferr}
We observe that the spin Hamiltonian we consider, both for deformed and undeformed cases, is invariant under the symmetry transform.

$$\Theta=\hat R_z(\pi)\mathcal{T},$$
where 
$$\hat R_z(\pi)=e^{-i \pi \sum_i S_i^z}$$
is a 180-degree rotation of each spin around the z-axis,
$$\mathcal{T}= e^{-i \pi \sum_i S_i^y} K$$
a time reversal transformation that consists of a 180-degree rotation around the y-axis and complex conjugation $K$.
The invariance of the Hamiltonian becomes clear if we observe that spin operators transform as
$$\Theta S_i^x\Theta^{-1}=S_i^x; \quad \Theta S_i^y\Theta^{-1}=S_i^y;\quad \Theta S_i^z\Theta^{-1}=-S_i^z$$
This transformation not only leaves the Hamiltonian invariant but also, very crucially, the anti-ferromagnetic ground state that is restricted to the $x$-$y$-plane. Therefore, the corresponding magnon Hamiltonian will also be left invariant.

One may now expand using a Holstein Primakoff expansion to linear spinwave order (expansion around the classical ground state) to obtain
$$\Theta=e^{-i \pi N S} \exp \left[i \pi \sum_i a_i^{\dagger} a_i\right] \exp \left[-\pi \sqrt{\frac{S}{2}} \sum_i\left(a_i-a_i^{\dagger}\right)\right] K.$$
Applying the transform to creation and annihilation operators, we obtain (again to linear spinwave order)
$$\Theta a_i \Theta^{-1}=a_i^{\dagger}, \quad \Theta a_i^{\dagger} \Theta^{-1}=a_i.$$
It is now easy to see that the Nambu spinor
$$
\boldsymbol\Psi(\vec k)=(\vec a(\vec k),\vec a^\dag(\vec k));\quad \vec a(\vec k)=\sum_{i}e^{i\vec k\cdot \vec r_i}\vec a_i$$
transforms as
$$\Theta\boldsymbol\Psi(\vec k)\Theta^{-1}=\tau_x\boldsymbol\Psi(-\vec k);\quad \tau_x=\left(\begin{array}{cc}
0 & \mathbb{1} \\
\mathbb{1} & 0
\end{array}\right)$$
From the invariance of the many-body Hamiltonian under the symmetry transform, it follows for the single-body Hamiltonian
$$h(\vec{k})=\tau_x h(-\vec{k})^* \tau_x .$$
and similarly the dynamical matrix $D(\vec k)=\eta h(\vec k)$ now fulfill
$$D(\mathbf{k})=-\tau_x D^*(-\mathbf{k}) \tau_x.$$

It then follows from conjugating the eigenvalue condition and multiplying by $\tau_x$ that in a suitable gauge
$$\left|u_n(-\mathbf{k})\right\rangle=\tau_x\left|u_n(\mathbf{k})\right\rangle^*.$$
We stress that, here, we assumed we are dealing with an isolated band, consistent with bands for which the Chern number is well-defined.

For the Berry connection, then follows that
$$\mathbf{\mathcal{A}}_n(\mathbf{k})=-i\eta_{nn}\bra{u_{\mathbf{k}}}\eta \nabla_{\mathbf{k}}\ket{u_{\mathbf{k}}}=-\mathbf{\mathcal{A}}_n(-\mathbf{k}),$$
where we used $\tau_x\eta\tau_x=-\eta$. Now since the Berry curvature is given as $\Omega(\mathbf{k})=\nabla_{\mathbf{k}} \times \mathcal{A}(\mathbf{k}),$
It immediately follows that
$$\Omega(\mathbf{k})=-\Omega(-\mathbf{k})$$
is antisymmetric. 
\end{document}